# Waveform cross correlation for seismic monitoring of underground nuclear explosions. Part II: Synthetic master events.

Dmitry Bobrov, Ivan Kitov, and Mikhail Rozhkov

Abstract

Waveform cross correlation is an efficient tool for detection and characterization of seismic signals. The efficiency critically depends on the availability of master events. For the purposes of the Comprehensive Nuclear-Test-Ban Treaty, cross correlation can globally reduce the threshold monitoring by 0.3 to 0.4 magnitude units. In seismically active regions, the optimal choice of master events is straightforward. There are two approaches to populate the global grid in aseismic areas: the replication of real masters and synthetic seismograms calculated for seismic arrays of the International Monitoring System. Synthetic templates depend on the accuracy of shape and amplitude predictions controlled by focal depth and mechanism, source function, velocity structure and attenuation along the master/station path. As in Part I, we test three focal mechanisms (explosion, thrust fault, and actual Harvard CMT solution for one of the April 11, 2012 Sumatera aftershocks) and two velocity structures (ak135 and CRUST 2.0). Sixteen synthetic master events were distributed over a $1^o$x$1^o$ grid. We built five cross correlation standard event lists (XSEL) and compared detections and events with those built using the real and grand master events as well as with the Reviewed Event Bulletin of the International Data Centre. The XSELs built using the source of explosion and ak135 and the reverse fault with isotropic radiation pattern demonstrate the performance similar to that of the real and grand masters. Therefore, it is possible to cover all aseismic areas with synthetic masters without significant loss in seismic monitoring capabilities based on cross correlation.

# Introduction

In Part I of this paper, we have presented a thorough analysis of grand master events as a workhorse for the waveform cross correlation (CC) techniques developed at the International Data Centre (IDC) for global seismic monitoring. A grand master is a real event with the best quality signals measured at several IMS arrays stations. We assumed that focal mechanisms and source functions are generally similar within the same seismic zone, and thus, it is possible to select one real event, which is the most effective for cross correlation with all events in a relatively big footprint of the selected event. The use as a grand master implies that this real event is replicated over a regular grid. The original signals at individual sensors of the involved arrays are retained and the arrival times for the grid positions are shifted by the relevant theoretical delays. For the replicated grand master templates, all improvements in detection, phase association, and event building obtained with real master events in their actual positions can be extended by hundreds of kilometers beyond the boundaries of seismically active areas. For seismic monitoring purposes, Bobrov *et al.* (2012ab) showed that cross correlation with real master templates allows finding 50% to 100% events in addition to those reported in the official Reviewed Event Bulletin (REB), all matching the full set of event definition criteria (EDC) currently adopted by the IDC. These added events are generally of lower magnitudes and reduce the level of threshold monitoring (Kværna *et al*., 2007) by 0.3 to 0.4 magnitude units. The detection threshold in immediate proximity to the best master events can be reduced by 1.0 to 2.0 units of magnitude relative to that obtained by standard detection procedures used by the IDC (*e.g*., Schaff *et al*., 2012). The replicated grand masters moved hundreds of kilometers from their actual positions demonstrate no significant loss in the magnitude resolution and detection sensitivity.

Formally, one can cover the whole globe with a few grand master events and build up a global grid (GG) for comprehensive seismic monitoring using waveform cross correlation. In practical terms, such a global grid is the principal goal of our research. However, the underlying advantages of cross correlation associated with similar mechanisms, source functions, propagation paths, and empirical travel time delays at individual sensors of the IMS arrays stations degrade with the master/slave distance. Therefore, concerning the global coverage, the empirical grand master events compete with synthetic waveforms modeled using seismological and geophysical information available from all sources. These sources span the whole range of global and regional velocity/attenuation models, including the most detailed models of the crust, focal mechanisms and source functions likely associated with tectonic stresses in aseismic areas, and fine velocity structures beneath the IMS array stations. The data and models are accompanied by extended experience with modeling of real earthquakes and historical nuclear test explosions. The latter events are of special interest for seismic monitoring under the CTBT and deserve a deeper study using the entire set of underground nuclear explosions (UNE) measured worldwide by arrays and 3-C stations.

The idea of using synthetic master events was naturally born from the need of covering the globe with masters distributed over a grid of certain density, taking into account the fact that vast areas of the earth are aseismic and real seismic events cannot be used as masters in aseismic areas. Rogers *et al*. (2006) introduced cross correlation with model-based signals as an advanced monitoring method for aseismic areas. They used a set of 3D velocity models to calculate intermediate period (>10s) seismograms for a moderate (Mw~5) event at the China-North Korea border. The obtained synthetic signals showed an excellent level of fit and high cross correlation coefficients at a number of regional stations. However, the underlying 3D velocity models require an exceptional computation power for short-period signals at far-regional and teleseismic distances, which are most important for global monitoring of low-yield UNEs. Here, we make a step to the global application of the cross correlation technique in seismic monitoring by using short-period synthetic seismograms at teleseismic and regional distances as master templates. By comparison with the results obtained using real (Bobrov *et al.*, 2012b) and grand masters (Part I of this paper), we demonstrate that cross correlation retains the level of performance appropriate for effective detection and association of seismic signals when synthetic seismograms are calculated in 1D velocity models and for simple focal mechanisms. Therefore, one does not need sophisticated 3D models and accurate focal mechanisms to populate the aseismic part of earth with synthetic master events.

In order to assess the performance of model-based signals relative to real and grand masters we generate a set of synthetic templates at seven primary IMS array station for a regular grid of master events covering the aftershock zone of the 2012 Sumatra (Ms(IDC)=8.6) earthquake. Theoretical seismograms take into account the velocity/attenuation structure between source and receiver, including the 2°x2° crust velocity model, and a variety of source mechanisms. Then we cross correlate continuous waveforms at the involved IMS stations with various synthetic master templates and create a series of detection lists for the same time interval as was used for the real and grand masters. After phase association, event building, and conflict resolution we build cross correlation standard event lists (XSEL), which include only the events meeting the EDC.

## Synthetic seismograms

Building full wave-field synthetic seismograms is a time consuming computational process and experimenting with huge amount of data demands extensive human efforts and computer resources. Since the CC master event technique involves teleseismic P-waves and far-regional Pn-waves, a simplified approach can be used based on seismic inversion technique introduced by Hudson (1969a) and developed later by Robert Herrmann (*hudson96* program of his comprehensive software package "Computer Programs in Seismology", CPS). Hudson's work can be considered as an improvement of Carpenter's (1966) attempt to construct realistic pulse shapes for the teleseismic body waves from underground explosions and providing the solutions for underwater and atmospheric explosions as well. He developed the algorithm to calculate body and surface waveforms at teleseismic distances in a layered elastic half-space (waves recorded at epicentral distances between 30° and 100°) from both explosive and shallow earthquake sources. Hudson considered the body waves travelling through the mantle without being affected by the core before arriving at the mantle/crust boundary below the receiver. The approximation is done for teleseismic distances so the wave fronts are spheres centered on the point R=0 at the base of the layers vertically beneath the source. This approximation can be used to describe body waves propagating through the mantle if the separation of the rays at the base of the crust is small compared with the epicentral distance of the receiver. Carpenter (1966) showed that rays at epicentral distances between 30° and 95° have angles of incidence between 7° and 14° at the free surface. Then, the separation of a few kilometers between the rays at the base of the crust may be neglected (Hudson, 1969b).

The *hudson96* code uses the Haskell's matrix method to account for layering at the source and receiver. Attenuation due to linear anelasticity is allowed by an empirical factor, and the sphericity of the earth is accounted for by a geometrical spreading factor. This code solves the problem in the frequency domain and uses geometric-ray theory for propagation of body waves through the mantle (Thigpen, 1979). The *hudson96* program is aimed at the direct teleseismic P or S arrivals using a stationary phase approximation (Aki and Richards, 2002) for the ray parameter connecting the direct arrival from the source to the receiver. This approach uses the propagator matrices for the source and receiver crust structures, so that the depth phases are automatically included as well as the details of the source and receiver crustal velocity models. The computation is fast because of the assumption that only one ray parameter is required. This assumption is fine for shallow earthquakes for which the P, pP and sP have the same ray parameter, but may not be appropriate for large epicentral distances and 700 km depths (Herrmann, CPS documentation). The travel time, t*, and geometrical spreading are computed within the program. However, t* can be adjusted by a user which can be useful for poorly defined models. We tested different models with t* varying in a range 0.5-1.2 in order to provide adequate spectral content of synthetic records.

A comparison of results of *hudson96* teleseismic P-wave synthetics with full wavenumber integration technique for the ak135 velocity model was analyzed by Herrmann (2006). This comparison was made for the following focal mechanism: 45° dip-slip, 90° dip-slip, vertical strike-slip, and for an explosion source. It was shown that the P, pP, and sP segments of the *hudson96* synthetics agree well with the complete synthetics, and the slight difference might be associated with the sampling granularity.

In our modeling, we created synthetic seismograms for a wide range of far-regional and teleseismic distances (Δ from 17° to 81°) for seven IMS stations involved in the previous cross correlation study of the Sumatera aftershocks. We used the ak135 model and built a set of local source and receiver models using CRUST 2.0 project data. Despite *hudson96* program is designed to build synthetic seismograms at teleseismic distances, it showed appropriate results even for far-regional distances (in particular, for station CMAR at about 17°) in terms of cross correlation. We had also tried other modeling approaches before we have chosen the stationary phase approximation approach: the Cagniard de-Hoop method and wavenumber integration. We came to the same conclusions as Herrmann in CPS documentation.

Without loss of generality, a few types of sources were tested in order to assess the performance of synthetic masters. It is clearly understood that, in the areas where real masters cannot be obtained, the choice of appropriate waveform templates is a challenge. A promising opportunity consists in synthesizing a representative master accurately describing the distinctive features associated with a certain class of seismic events. There is a simplified approach and a sophisticated one. The latter may include the use of subspace detectors (Harris, 2006; Harris and Paik, 2006; Harris and Dodge, 2011) for a multitude of synthetic waveforms, comprising the events with varying source time history, focal mechanism, depth, and near-source velocity model. This is a realistic task but it requires time and resources not available at the stage of feasibility study. Here, we compare the performance of modeled waveforms generated by simple quasi-omnidirectional and isotropic sources (*i.e.*, all generated waves have positive first arrival polarity) with that demonstrated by the synthetic seismograms calculated for one event with known CMT solution.

The general force-couples for a deviatoric moment tensor can be represented by three fundamental faults: a vertical strike-slip, a vertical dip-slip, and a 45° dip-slip (Dreger, 2003; Herrmann, 2006). The base of our first representative source is a simple 45° degree dip-slip fault model assuming the following focal mechanism: strike $\theta=0°$, dip $\delta=45°$, and slip $\lambda=90°$ (thrust faulting). The upper row in Figure 1 displays the focal mechanism, style of faulting, and radiation pattern. Obviously, real sources are more complicated than this ideal thrust fault. In order to approximate lower variations in actual radiation patterns from the observed aftershocks we have extended this standard model by creating a quasi-omnidirectional dip-slip source. This is the same 45° dip-slip source with the isotropic radiation, i.e. the source having uniform seismic energy radiation for all azimuths. Further, this source is called "Standard Isotropic" or "STI".

Our second representative is the thrust fault ($\delta=45°$, $\theta=90°$, and $\lambda=90°$) represented in the lower row of Figure 1. We choose this type of source due to its radiation pattern providing preferential isotropy for the IMS stations participating in the analysis. This source is called "Standard", or "ST". For both sources in Figure 1, we used ak135 model for the propagation in the mantle and CRUST2.0 to represent crustal velocity structures below the source and receiver. Therefore, these are effectively 2D velocity models. The next two cases under consideration are based on the same explosion source function generating synthetic seismograms in different velocity structures: (a) 1D model (ak135) called "Expl1D"; (b) 2D models consisting of ak135 and CRUST2.0 called "Expl2D".

Synthetic seismograms at seven IMS array stations at ranges between 17° and 81° from the same location (1.5°N, 92°E) and depth of 10 km for all four sources are depicted in Figure 2. The involved IMS stations are ASAR, CMAR, GERES, MKAR, SONM, WRA, and ZALV. There are significant differences in the shape of the modeled signals, which should be manifested in cross correlation with continuous waveforms. All synthetic records have a feature of great importance for cross correlation – there is no ambient microseismic noise except very low rounding error associated with discrete calculations. In practice, the signal-to-noise ratio for synthetic templates is extremely large.

The fifth source is represented in the REB and in the Global Centroid Moment Tensor database (201204131509A). According to the IDC solution, this aftershock (orid=8605401) occurred on April 13, 2012 at 15:09:21.00 GMT, was located 0.9886°N, 92.140°E, and, by IDC rules associated with the depth uncertainty, was fixed to zero depth. The CMT solution is as follows: date: 2012/04/13; centroid time: 15:09:25.8; 1.01°N, 92.16°E; D=12.0 km; Mw=4.9; mb=4.3; the used fault plane: $\theta=117°$, $\delta=47°$, $\lambda=174°$. Figure 3 displays the double-couple representation of this source, and the upper panel of Figure 4 compares seven synthetic seismograms obtained for the IDC location at the involved IMS stations. All synthetic seismograms are convolved with the relevant instrument responses and the measured waveforms are taken from the channels providing visually good signals. We have band-filtered all signals between 0.6 Hz and 4.5 Hz in order to improve the signal-to-noise ratios. In Figure 4, all records are scaled to their peak values. This filtering is necessary since the originally measured traces are heavily contaminated by the ambient low-frequency noise. Overall, the synthetics are similar to the observed signals.

We intentionally avoid any velocity model/source tuning, which is practically a mandatory part of seismic wavefield modeling. Under the cross correlation framework, it is assumed that synthetic masters will cover vast aseismic zones, where no model/source adjustment is possible using actual observations. One of the problems with the theoretical CMT solution, which is obtained using the P-waves of longer periods, consists in a stronger variation of modeled amplitudes than the observed ones. The lower panel of Figure 4 compares the observed and simulated spectra at three IMS stations: CMAR, MKAR, and SONM, the upper curves correspond to synthetics. For cross correlation with synthetic seismograms, we use the same time windows as for real master events (Bobrov *et al*., 2012c). Therefore, only ten seconds after the first arrival are necessary. For an array, we use the same record at all individual sensors with time delays associated with a given source/station configuration. At the same time, actual waveforms differ from sensor to sensor and actual time delays differ from the theoretical ones. These differences suppress cross correlation between real and synthetic signal. Fortunately, this suppression is similar for valid signals and the ambient noise and the level of signal-to-noise ratio suffers much less than cross correlation coefficient.

## Waveform cross correlation

In order to assess the relative performance of real and synthetic master events we have carried out calculations of cross correlation in various settings for the aftershock sequence of the Sumatra 2012 events. Bobrov *et al.* (2012c) presented an extensive analysis of cross correlation using real master events in their actual positions. In Part I, we put one strong grand master event (SGM) in the actual positions of the real master events and also distributed it over a regular grid with $1^{o}$ spacing, as shown in Figure 5. In Part II, we calculate synthetics for seven IMS stations and one source placed in sixteen nodes of the same grid and process 24 hours after the main shock using the same set of procedures as for the real and grand masters. There are five different synthetic events and thus five different detection and event lists to compare with the previous results.

The full set of procedures associated with cross correlation is described in Part I. Here, we highlight the most important features. The cross correlation coefficient, *CC*, is calculated using waveform templates, which include several seconds of (P- or Pn-wave) signal and a short time interval before the signal. To enhance the detection of weak signals continuous waveforms are processed in four frequency bands in parallel. For the low-frequency (BP, order 3) filter between 0.8 Hz and 2.0 Hz, the length is 6.5 s and includes 1 s before the arrival time. For the high-frequency filter between 3 Hz and 6 Hz, the full length is only 4.5 s. For Pn-waves, the length is 11 s and does not depend on frequency. The band with the best quality cross correlation signal is chosen to calculate its attributes.

For array stations, we calculate *CC* channel-by-channel and then average individual traces to obtain a single *CC* time series (Gibbons and Ringdal, 2006, 2012). This trace is used to detect signals with standard short-term/long-term moving average (STA/LTA), which is also considered as the signal-to-noise ratio characterizing the CC-trace, *SNR_CC*, and to estimate the peak *CC* for these signals. The length of the short-term and long-term window is 0.8 s and 40 s, respectively. For synthetic templates, we retain the detection threshold *SNR_CC*>2.8 used and the previous studies and reduce the |*CC*| threshold to 0.1, i.e. no detection with |*CC*| <0.1 is possible. Both thresholds have to balance the rate of missed valid signals and false alarms.

Individual *CC*-traces provide reliable azimuth and slowness estimates using standard *f-k* analysis. These parameters define the quality of detections together with *CC* and *SNR_CC*. To link the CC-based and current IDC processing, for all signals detected by cross correlation standard attributes are estimated: amplitude, period, azimuth, scalar slowness, SNR, quality class, etc. Where applicable, we calculate body wave magnitude, which is also used in global association of seismic phases. Since higher cross correlation coefficients generally indicate the spatial closeness between master and slave, Bobrov *et al*. (2012b) proposed to use $|x|/|y|$, where *x* and *y* are the vectors of data for the slave and master event, respectively, as a measure of relative size. The logarithm of this ratio, $RM = \log(|x|/|y|) = \log|x| - \log|y|$, is the magnitude difference or the relative magnitude, *RM*.

Then we try to associate the arrivals obtained by a given master and to build events. In this procedure, we consider only the signal attributes estimated using cross correlation. Unlike the Global Association currently used at the IDC (Coyne *et al*., 2012), which is based on a sophisticated algorithm, the cross-correlation based association is much simpler. Since we are looking for close events the association of phases is just local, *LA*. At a given station, the travel time difference between a master and a slave cannot be more than, say, 6 seconds. The master/station travel times can be used to project the measured arrival times back to source. As a result, a set of origin times at IMS stations is obtained. To build a viable REB event hypothesis, one needs three or more origin times at different stations to group within, say, 8 seconds; all attributes of the associated signals (azimuth, slowness, relative magnitude) have to be within the station-specific uncertainty bounds (Coyne *et al*., 2012).

Apparently, the global grid of master events is coarse (1° spacing) for location purposes. To account for the master/slave distance, we introduced two concentric rings consisting of 6 and 12 virtual masters at distances 0.225° and 0.450°, respectively. Figure 5 illustrates the distribution of these virtual masters for all sixteen nodes. Theoretical travel times are recalculated for all virtual masters and then used to re-estimate the relevant origin times for the local association. So, we build 19 alternative hypotheses and select the one with the largest number of stations and the lowermost origin time RMS residual.

After resolving a number of conflicts between attributes of the arrivals in the same hypothesis we obtain a set of REB-consistent hypotheses for individual master events.

However, adjacent masters may find physically the same event. To avoid redundancy, we again chose the competing events the one with the largest number of stations and the smallest origin time RMS residual. All detections associated with the unsuccessful hypothesis are eliminated. When all conflicts between the hypotheses associated with different masters in a given time interval are resolved we obtain an XSEL.

Any qualified event in the XSEL should formally meet the EDC but can also be characterized by several parameters related to cross correlation. In the *LA* process, the relative station magnitudes are determined and then averaged to estimate the network relative magnitude *RM*. Bobrov *et al.* (2012b) showed that the standard deviation of the network *RM* is much lower than that of body wave magnitude. The average cross correlation coefficient, *CC_AVE*, characterize the average quality of associated arrivals and the cumulative one, *CC_CUM*, can be interpreted as event quality.

## Comparison with real and grand masters

We have to estimate the performance of five synthetic templates relative to those generated by the real master events and grand masters. These templates are synthetic seismograms calculated for the reference elements of seven IMS array stations and correspond to 16 nodes of the regular grid shown in Figure 5. It should be proven statistically that the number and defining characteristics of detections (*e.g.*, *CC*, *SNR_CC*, *RM*) and events (*e.g.*, *CC_AVE*, *CC_CUM*) obtained by individual masters and those retained in the relevant XSELs after all conflicts between similar events are resolved have similar values and distributions.

There are five different versions of synthetic master events and it is instructive to select one or two best performers and then to compare them to the real and grand masters. The real master events (REAL) and two versions of the strong grand master (SGM) replicated over the locations of real masters (SGM_REAL) and over a regular grid (SGM_GRID) create three different cases for comparison with the best synthetic masters. We consider this regular grid as a part of the global grid. Therefore, when testing the performance of synthetic masters distributed over a small grid one actually assess the performance of the global grid and cross correlation as a monitoring technique.

In Part I, we have developed a procedure, which compares arrivals and events at all stages of processing – from detection to interactive analysis. In the order of processing, one obtains raw detections, the events built by the *LA* using individual masters and the arrivals associated with these events, the events left in the XSEL after conflict resolution with corresponding arrivals, and the events and arrivals matched in the Reviewed Event Bulletin extended by all REB ready events found by cross correlation. In this study, we use the official REB as a proxy to the XSEL after interactive analysis, which likely contains 50% to 100% more events matching the event definition criteria than the REB. In addition, the official REB may contain a few percent of spurious events. These are the events with wrongly associated phases from different sources or later phases of the same sources. Overall, the REB is not a perfect catalog to be repeated by XSEL.

Table 1 presents general statistics for five synthetic sources, the real masters (REAL), and two configurations of the SGM. We skip here the total number of detections, which slightly varies among the synthetic masters but does not influence the content of the relevant XSELs. The total number of associated detections is similar in all eight cases and varies from 55,651 for the standard source mechanism (ST) to 66,853 for the case of explosion in the 1D velocity structure (Expl1D). For the standard mechanism without azimuthal dependence (STI), the number of detections is similar to two explosion cases because the distribution of waveform amplitude is isotropic. The real masters detect a slightly lower number of signals than Expl1D, but more than other four synthetic cases. The SGM_REAL is the second best set of masters with 66,552 detections.

The number of events built by 16 masters also varies. As expected from the total number of associated phases, the Expl1D master creates the highest number of events (17,197). Slightly counter intuitively, fewer events (15,934) are built in the case of real masters. An explosion in the 2D velocity structure created more events (16,132) than the standard isotropic source (15,738) despite of the lower number of associated arrivals: 61,139 and 62,126, respectively. The standard source mechanism is inferior to all other sources in the number of created events and associated phases.

After the conflict resolution, the case Expl1D retains its lead among synthetic masters in the number of events (1,812) and associated phases (7,489). The real masters outperform all other masters in the number of events (2,324) and associated arrivals (9,631). This is the only setting where all 16 masters have different focal mechanisms, actual depths, magnitudes, and time functions. The templates generated by 16 real masters should better represent the whole diversity of waveforms observed from the aftershocks than any individual, especially synthetic, master. A major deficiency of all synthetic masters and simple velocity/attenuation models is the failure to accurately predict the observed variation of signal amplitudes over IMS stations. As shown below, the relative template amplitudes at seven IMS stations are biased for all synthetic masters and the relative magnitude, *RM*, at these stations fluctuates in a wider range than for real aftershocks. Therefore, the scattering of the difference between station and network average RM is higher and extends beyond the predefined uncertainty limit $|dRM|<0.7$. As a result, we miss many valid arrivals due to their artificial dynamic inconsistency. When the d*RM* threshold is removed from the LA processing ("No RM"), the number of built events and associated arrivals in Table 1 increases.

The overall detection and event statistics might be slightly confusing. The defining characteristics of the detections and events may vary depending on the source mechanism of synthetic master and velocity structure. It is instructive to compare the detailed distributions of defining parameters for five synthetic masters. Figure 6 illustrates the quality of the events built by the studied masters before and after conflict resolution as expressed by the probability density functions (PDF) of the event average *CC, CC_AVE*. Except two outliers, all curves in both panels are practically identical. The CMT solution demonstrates relatively fewer events with higher *CC_AVE* values, and the STI source mechanism lacks events with low average *CC*. The conflict resolution removes all events with the highest average *CC*. This effect is associated with the number of stations as the primary defining parameter. The events with the highest *CC_AVE* are likely missing the stations with low *CC*, and thus, are rejected by the conflict resolution algorithm.

Figure 7 displays the PDFs for cumulative *CC, CC_CUM*, i.e. the average *CC* times the number of stations in a given event, before and after the conflict resolution. The original distributions are counted in 0.3-wide bins starting from 0.6. This is the lowermost possible *CC_CUM* for an event defined by three stations with |*CC*|=0.2. The bins are centered at *CC_CUM* 0.75, 1.05, … For the whole sets of events built by 16 masters, all curves practically coincide between 0.75 and 2.35 and peak at *CC_CUM*=1.05. Beyond 2.55, the CMT curve falls faster than four other PDFs, and the Expl1D curve is the second best after the S curve. For the XSEL events shown in the right panel of Figure 7, the PDFs are much closer and diverge only for the highest *CC_CUM*, where the number of events defining the PDFs is negligible. One may conclude that all five synthetic masters have practically the same resolution in terms of the average and cumulative *CC* in the XSEL.

The PDFs of relative magnitude depicted in Figure 8 depend on the accuracy of station amplitude prediction by five source mechanisms. Two mechanisms (CMT and ST) have azimuthal dependence of seismic wave amplitude and three are isotropic. The shapes of all five curves are similar, but the STI mechanism overestimates station amplitudes by a factor of 30 (1.5 units of magnitude) while the CMT source underestimates the station magnitudes by ~1.0 unit. These biases are induced by source calibration problems and are easily removed. It is important that calibration of synthetic seismograms does not affect the performance of master events as expressed by the found arrivals and built events. All synthetic signals have extremely high SNRs defined only by the truncation error.

The quality of arrivals found by cross correlation can be characterized by *CC* and *SNR_CC*. The level of *CC* is a good characteristic of quality at a given station since it varies between 0 and 1.0 and thus differs from amplitude of original signals. For an array with an aperture of 20 km and two dozen individual sensors, beaming of cross correlation traces to a master event is a very effective measure of noise suppression. Then signals with |*CC*| ~0.1 may have *SNR_CC*> 2.8. Small aperture arrays with ten or fewer sensors are less effective in noise suppression in the CC traces. As a result, a significant portion of detections with |*CC*| ~0.3 have signal-to-noise ratio below the threshold of 2.5 adapted in our study. In Figures 9 and 10, we present the PDFs of *CC* and *SNR_CC* estimated for five synthetic masters from all associated phases and from those in the relevant XSELs. The CMT curve has the largest portion at the highest CCs and SNR_CCs. The STI curve demonstrates a higher density between *CC* 0.4 and 0.6, but falls faster at lower and highest CCs. Five SNR_CC curves are hardly separated between 2.5 and 6.0 for all associated phases and fluctuate above 8.0, where the number of detections is rather negligible and thus unrepresentative. In the right panel of Figure 10, we compare two PDFs estimated for the XSELs for the same synthetic master (Expl1D) with and without the d*RM* threshold of 0.7. Between 2.5 and 6.0, there is no difference between these two cases. The NoRM PDF is marginally higher between 6.0 and 8.0. From Figures 9 and 10, one may conclude that all synthetic masters create similar distributions of CCs and SNR_CCs. In that sense, they are interchangeable.

Figure 11 depicts PDFs of the station deviation from the network relative magnitude, d*RM*. These PDFs are calculated from the XSELs generated by five synthetic masters with and without dRM threshold applied. In the left panel, the largest possible |d*RM*| is 0.7 and all PDFs are confined to this range. The STI curve is characterized by the narrowest PDF range between 0.04 and 0.09, and is centered on d*RM*=0.0. Other curves are slightly shifted from the center (between 0.1 and 0.2) but have a wider dynamic range (from 0.03 to 0.12) with a prominent fall at larger |d*RM*|. All five curves in the left panel have thick tails, which suggest that some good arrivals might be missed due to the dRM threshold. When this threshold is abandoned, the PDFs for the XSEL arrivals are extended and well above zero to |d*RM*|~1.0 to 1.5 as the right panel shows. The observed change in the overall shape and peak values (e.g., STI and Expl1D) implies that the NoRM XSELs have different sets of associated phases. Obviously, there are additional arrivals with |d*RM*|>0.7, but there are also many new arrivals with |d*RM*|<0.7 because new events are built. In the conflict resolution process, these new events may have more defining stations and/or cumulative CC than competing XSEL events built with the RM threshold. The former events should replace the latter ones.

Our estimates show that, for regions without real seismic events, synthetic masters are not able to accurately predict amplitudes at IMS stations. There are two ways to address this observation in the LA process: to extend the dRM range or abandon it at all. We used the latter approach in this study. The relative magnitude can be effectively used for event building, however, when real masters are available.

We have compared the overall performance of five different synthetic masters in terms of detection, phase characterization/association, and conflict resolution. For the sake of simplicity, we retain two cases for further analysis: Expl1D better represents the source of explosion and the STI solution is a natural choice to characterize earthquake seismograms with an isotropic source function. This is not our current task to solve the inverse problem for the entire variety of source mechanisms associated with the Sumatra's aftershock sequence and to determine the best ones. Synthetic masters are designed to cover the globe in aseismic areas, where no earthquake mechanisms are available. We also retain only the XSELs obtained without RM threshold.

In Part I, we presented five sets of actual master events. Sixteen actual master events (REAL) selected from approximately 1,000 Sumatra aftershocks provided the largest number of XSEL events. One of these actual masters was replicated over the position of 16 actual masters (SGM_REAL) and over a regular grid with a 1° spacing (SGM_GRID). Both replicated master sets gave results similar to the real set. Figure 12 compares various characteristics of the detections in the XSELs obtained using real and synthetic masters. The PDFs of two synthetic masters are lower for |CC|>0.35. However, all SNR_CC curves are similar between 2.5 and 8.0. Therefore, the level of CC associated with two synthetic masters falls proportionally for the detected signals and the ambient noise. Accordingly, the number and the quality of arrivals obtained using the synthetic and real masters do not change. The STI dRM curve in Figure 12 differs from those for real masters. The Expl1D PDF, however, is very similar to three real PDFs, but has a slight positive bias, while the real curves are characterized by a negative bias. This problem is resolved by amplitude calibration of synthetic seismograms, where available.

Bobrov *et al.* (2012b) described the process how multichannel traces of cross correlation coefficient are used to estimate (pseudo-) azimuth and (pseudo-) slowness using standard *f-k* analysis. For array stations, azimuth and slowness residuals are effective filters rejecting all arrivals from directions and distances different from those related to a given master. For all arrays, we defined the limits of uncertainty as 20° and 2 s/deg, respectively. Apparently, these limits should depend on station aperture and we will demonstrate the difference in station resolution later on.

Figure 12 depicts five overall PDFs for azimuth, d*AZ*, and slowness, d*SLO*, residuals. The PDFs associated with real masters peak at azimuth and slowness slightly deviating from zero. Both REAL curves related to 16 different real masters are sharper than others, especially for d*AZ*. For a given event, the deviation between theoretical and actually observed azimuth and slowness is compensated by the slowness-azimuth-station-corrections (SASCs), which are pre-calculated for many seismic areas. These deviations are induced by non-planar propagation of seismic wave fronts across arrays. Since *f-k* analysis seeks for a plane solution with the maximum weighted coherency between all channels the deviations from the theoretical arrival times at individual sensors create a virtual shift in vector slowness. SASC have to partially compensate the overall shift and both residuals are calculated relative to the SASC corrected values, i.e. relative to actually observed azimuth and slowness for a given master. Therefore, real masters with SASCs create symmetric distributions centered near zero. The synthetic masters do not include SASCs, i.e. we use theoretical time delays relative to reference channel of a given station. In Figure 12, both synthetic masters peak at azimuth -5° and slowness -0.8 s/deg. These overall distributions include all 7 IMS stations with varying number of arrivals. When all arrivals are corrected for the relevant SASCs, the d*AZ* and d*SLO* distributions for both synthetic masters would be centered at zero values. Then, one could have associated more arrivals found by cross correlation. For an aseismic region without SASCs, it could be helpful to introduce a wider range of allowed azimuth and slowness residuals or to estimate SASCs using different approach. Having time delays at individual sensors from seismic waves with various azimuths and slownesses, one might recover the 3D velocity structures under all IMS arrays using seismic tomography. Then SASCs will be calculated for any azimuth and slowness.

We have studied a variety of agregate distributions. On average, the real and synthetic events and associated phases behave in a similar way with a few important biases caused by the difference between real and modeled array seismograms. At individual stations, these biases might be prominent. When known, even large static errors can be easily compensated with a signifcant improvement in the performace (i.e., the number of detections and events) and statistics of cross correlation techniques.

Table 2 presents general statistics of station arrivals in the relevant XSELs for three real and five synthetic sets of masters. The latter ones have two versions − with and without the dRM threshold. Station ASAR demonstartes similar figures for all cases with a slight deficit for both SGM sets. For CMAR, the set of 16 real masters is the most efficient in terms of associated detections with the NoRM Expl1D as the second best master. At GERES, REAL is also the best but the Expl2D with the dRM threshold marginally outperforming both versions of Expl1D. Stations MKAR, WRA, and ZALV show similar behavior of synthetic and real masters: REAL is always the best and there is no set which performs much worse than the others. Station SONM reveals a dramatic change when dRM is applied. For the CMT solution, there is almost no associated phases (15). This feature implies an extremely large bias in amplitude prediction at SONM relative to real signals and other stations. This observation practically prohibits the use of the CMT solution as a synthetic master. The STI is likely the best master for SONM − the number of associated arrivals does not fall too much when dRM is used. Overall, the Expl1D and STI masters perform at the level of both SGM sets and thus can be used for the purposes of monitoring based on waveform cross correlation.

We start the assessment of specific parameters with the relative magnitude. For seven IMS array stations, Figure 13 compares the d$RM$ frequency distributions assocaited two sets of real masters (REAL and SGM_REAL) and two synthetic masters (Expl1D and STI) replicated over the grid. In Part I, we have already reported station biases in d$RM$ for real masters and discussed correction for static errors as a remedy. Both synthetic masters reveal even higher biases. For ASAR, the REAL curve is shifted by +0.1 to +0.2 and the SGM curve is centered at -0.1. The Expl1D frequency distribution is charaterized by +0.15 static error and falls to zero master than the the REAL curve. For STI, the distribution is shifted by -0.25. When corrected for the relevant biases, the Expl1D and STI curves for ASAR match the dRM threshold.

For CMAR, all curves are shifted by 0.2 (REAL) to 0.55 (STI). These shifts may result in a progressive loss of valid arrivals with |d$RM$|>0.7 in the relevant XSELs. Since all curves are symmetric, they are easily corrected for static errors and their peaks centered at d$RM$=0. The larger corrections will significantly change the network average for the XSEL events with CMAR as a defining station and also give birth to new events with new CMAR arrivals. This effect is most prominent for SONM. The STI curve is shifted by -0.55 and the Expl1D frequency distribution is centered at d$RM$=-0.75. Roughly, only half of arrivals obtained using the Expl1D master set meet the dRM criterion, as Table 2 shows. For the CMT master set, the RM bias at SONM is -1.55 and only 15 from 1051 arrivals meeting other criteria (e.g., *CC*, *SNR_CC*, origin time, azimuth, and slowness) can be associated with XSEL events according to the dRM rule. The trade-off between the d*RM* corrections and varying numbers of associated arrivals requires an iterative and network-wide approach. For a given set of masters, all curves for individual stations (one may extend the list of IMS stations for the studied aftershock sequence to 17) should be as accurately centered at d*RM*=0 as possible.

Figures 14 and 15 demonstrate the station-to-station variations in (pseudo-) azimuth and slowness, which are estimated using *f-k* analysis. Array aperture plays the defining role in the unceratinty of these estimates. Larger arrays (e.g., WRA) have narrower distributions, while small-aperture arrays (e.g. GERES) are characterized by a higher scattering and broader distributions. For WRA, the d*AZ* and d*SLO* curves are sharp and symmetric. However, two curves associated with synthetic masters center at -5° and -0.7 s/deg, respectively. The real masters include SASCs and thus the relevant distributions of azimuth and slowness residuals are centered at zero. For CMAR, all azimuth distributions are wider than for WRA and their centers shifted by -10° to -15°. Only the STI slowness distribution has a sharp peak and the other three have broader plateaus with a few local peaks. Stations GERES and ZALV reveal the worst performance of azimuth and slowness, respectively. This makes the azimuth and slowness limits used in the study almost worthless for synthetic masters. Stations GERES, SONM, and ZALV illustrate the fact that not all IMS stations have SASCs yet. This is the reason why all curves for both real master sets are also shifted. When collectively corrected, all curves might be centered around zero but this would not make the distributions sharper. One needs arrays of larger aperture to improve detection capabilities and the XSEL.

As in Part I, we compare the synthetic XSELs with the REB for the same period. Table 3 lists the number of events and, for each station, the number of associated phases matched by an REB event within given uncertainty of arrival (±6 s) time. We consider two cases: the lowermost number of phases matched by REB arrivals is 2 and 3. There are also two versions of synthetic XSELs: with and without dRM threshold. The total number of REB events during the studied period is 514.

## Discussion

Overall, our results show that the synthetic templates modeled using a 1D-velocity structure and the source of explosion or that of earthquake with isotropic P-wave radiation pattern provide effective master events for CC-based detection, phase characterization, location, and magnitude determination, which are the main parts of event building. This is a crucial finding for the global grid creation. We are going to build a tentative GG with an approximately 1° spacing, which includes from 25,000 to 30,000 nodes. The basic version will likely include one master replicated over the entire grid. For monitoring purposes, the simplest source function Expl1D and ak135 is a natural choice for the basic version. In a 1D-velocity structure, synthetic seismograms depend on distance and practically each master/station pair has a different template. In a degenerate case, one can use one template for all distances, and thus, all stations. Here, the subspace detector approach based on the samples with gradually increasing distance might provide the optimal template. The difference between waveforms is then defined by the difference in station specific instrument responses. For identical waveform templates for all masters, one needs to calculate the channel CCs only once and average them with various time delays corresponding to the master/station relative positions. In a way, it is similar to steering in beamforming and saves computational time. This also facilitates the procedure of conflict resolution between adjacent masters in line with *f-k* analysis: the peak *CC* corresponds to the winning master as the peak azimuth and slowness characterize the arrival.

Introducing the subspace detector concept into the CC-based global location process may bring certain benefit in a wider sense. In the example given above (Figure 4) some pairs (synthetic and real records) demonstrate rather good correlation (for example, cross-correlation coefficient CC~0.8 for records filtered at 0.5-2Hz for stations ASAR, SONA and CMAR) while correlation for some other stations is low (CC~0.5 for MKAR and WRA). However, this modeling was based on the depth estimate taken from the CMT catalog. The REB gives zero depth for this event. Additional modeling was performed with the arbitrary depths, ended up with the best depth estimate of 1.2 km as presented in Figure 16.

For this certain source depth, the average correlation coefficient for most of stations is between 0.7 and 0.85, except for CMAR with CC about 0.6. Cross correaltion coefficient for other depths varies but still belongs to the interval 0.6 to 0.85. This modeling points to the fact that the automatic location based on the synthetic master event can be enhanced by designing proper synthetic masters. However correlation enhancement for one particular event cannot guaranty the same improvement for others. For aseismic areas, we cannot predict the probability of occurring events at one depth (or having certain focal mechanism) or another. The global improvement of results can be likely achieved by applying the subspace detector approach, which would take into account the whole multitude of possible cases of seismic event generation, including explosions.

As an alternative, a grand master can be used to populate the GG including the aseismic areas. However, synthetic seismograms provide a more flexible tool and may incorporate the whole set of seismological and geophysical information pertaining to specific areas. The grand master approach is likely more effective around the zones with historical seismicity. At ranges of a few hundreds of kilometers, the principal features of sources and propagation paths, and thus, the empirical time delays at array stations do not change much.

In the statistical assessment of synthetics, the same procedures as for the real and grand masters are fully retained. We compare the overall number of detections and event hypotheses as well as their principal characteristics in order to quantify the difference at the network and station level. Two simple source mechanisms and the ak135 velocity model gave practically the same number of detections and hypotheses as the grand master. The real masters in their actual positions are slightly more prolific, but are not superior in finding of the REB aftershocks. It is worth noting that there is some room to enhance the performance of synthetic masters in terms of cross correlation. For the particular case of the Sumatera aftershock sequence, the work on model/source adjustment is planned for the overall improvement of CC processing. The extremely high importance of the empirical time delays at the IMS array stations has been demonstrated. The absence of SASCs and actual observations from sources in aseismic areas heavily bias the azimuth and slowness estimates obtained using synthetic waveforms. Hence, it is mandatory to recover the fine local velocity structures beneath IMS stations using tomographic methods and the whole lot of local/regional/teleseismic body and surface phases.

A more specific task for the cross correlation technique arises from the purposes of seismic monitoring under the CTBT. The real and grand master events may be very efficient in finding earthquakes but fail to have a good traction with signals generated by small UNEs. This problem needs a deeper investigation with historical explosions as masters and slaves. At first, we are going to use three DPRK tests as masters for the Sumatera aftershocks and evaluate their performance relative to real aftershocks and synthetics. Then, all available records of UNEs at IMS and non-IMS arrays and 3-C stations will be cross correlated with appropriate time shifts between channels, where applicable. The resulted set of CCs will shed light on the similarity between explosion waveforms, their scattering, and possible outliers. All these features are important for synthesizing master events optimal for the CTBTO.

## Disclaimer

The contents of this publication are the sole responsibility of the authors and can in no way be taken to reflect the views of the European Union and the CTBTO Preparatory Commission.

Tables

Table 1. Total number of arrivals, associated arrivals, and events build by different sets of master events.

| Master | # Detections | | | # Events | | |
|---|---|---|---|---|---|---|
| | Total | After CR | | Total | After CR | |
| | | RM | No RM | | RM | No RM |
| Expl1D | 66853 | 7489 | 8225 | 17197 | 1812 | 1898 |
| Expl2D | 61139 | 7138 | 8027 | 16132 | 1759 | 1853 |
| CMT | 57341 | 6581 | 7910 | 15259 | 1621 | 1828 |
| Standard Isotropic (STI) | 62126 | 7038 | 7480 | 15738 | 1671 | 1745 |
| Standard (ST) | 55651 | 6512 | 7448 | 55651 | 1606 | 1731 |
| REAL | 63583 | 9631 | | 15934 | 2324 | |
| SGM_REAL | 66552 | 7182 | | 16516 | 1659 | |
| SGM_GRID | 64514 | 7198 | | 16101 | 1669 | |

Table 2. The number of arrivals at seven IMS station depending on master events.

|  |  | ASAR | CMAR | GERES | MKAR | SONM | WRA | ZALV |
|---|---|------|------|-------|------|------|-----|------|
| REAL |  | 1014 | 1524 | 939 | 1725 | 1498 | 1706 | 1225 |
| SGM_GRID |  | 877 | 1094 | 736 | 1247 | 1031 | 1316 | 897 |
| SGM_REAL |  | 871 | 1136 | 730 | 1245 | 1009 | 1319 | 872 |
| Expl1D | No RM | 1204 | 1378 | 737 | 1359 | 1169 | 1321 | 1057 |
| Expl1D | RM | 1183 | 1278 | 726 | 1357 | 643 | 1275 | 1027 |
| Expl2D | No RM | 1220 | 1343 | 721 | 1354 | 1131 | 1311 | 947 |
| Expl2D | RM | 1173 | 1195 | 744 | 1358 | 410 | 1319 | 939 |
| CMT | No RM | 1126 | 1332 | 716 | 1413 | 1051 | 1275 | 998 |
| CMT | RM | 1047 | 1262 | 724 | 1320 | 15 | 1202 | 1011 |
| STI | No RM | 1105 | 1252 | 698 | 1265 | 1099 | 1228 | 833 |
| STI | RM | 1107 | 1071 | 699 | 1242 | 924 | 1156 | 839 |
| ST | No RM | 1124 | 1294 | 691 | 1233 | 1084 | 1234 | 788 |
| ST | RM | 1062 | 1219 | 711 | 1219 | 317 | 1192 | 792 |

Table 3. For each of the involved IMS stations, the number of arrivals in the REB within 4 s from those in the XSELs.

|  |  | ASAR | | CMAR | | GERES | | MKAR | | SONM | | WRA | | ZALV | |
|---|---|---|---|---|---|---|---|---|---|---|---|---|---|---|---|
|  |  | **2** | **3** | **2** | **3** | **2** | **3** | **2** | **3** | **2** | **3** | **2** | **3** | **2** | **3** |
| **REAL** |  | 230 | 163 | 339 | 287 | 287 | 273 | 478 | 415 | 411 | 361 | 349 | 250 | 398 | 367 |
| **SGM_REAL** |  | 237 | 171 | 310 | 272 | 258 | 247 | 425 | 362 | 383 | 328 | 339 | 254 | 373 | 333 |
| **SGM_GRID** |  | 230 | 158 | 293 | 251 | 255 | 239 | 429 | 357 | 385 | 329 | 327 | 238 | 368 | 325 |
| **Expl1D** | RM | 227 | 133 | 246 | 212 | 254 | 227 | 388 | 315 | 182 | 147 | 295 | 172 | 350 | 300 |
|  | NoRM | 228 | 138 | 269 | 236 | 259 | 240 | 407 | 343 | 354 | 299 | 279 | 173 | 363 | 324 |
| **Expl2D** | RM | 235 | 143 | 218 | 183 | 259 | 232 | 391 | 304 | 105 | 87 | 294 | 185 | 352 | 294 |
|  | NoRM | 222 | 131 | 283 | 253 | 255 | 239 | 404 | 342 | 353 | 300 | 276 | 170 | 345 | 313 |
| **CMT** | RM | 216 | 127 | 273 | 226 | 253 | 226 | 379 | 286 | **2** | **2** | 294 | 177 | 372 | 293 |
|  | NoRM | 211 | 136 | 330 | 294 | 270 | 250 | 442 | 374 | 358 | 316 | 290 | 195 | 387 | 341 |
| **STI** | RM | 224 | 137 | 436 | 198 | 380 | 357 | 497 | 424 | 370 | 324 | 348 | 232 | 352 | 305 |
|  | NoRM | 220 | 140 | 456 | 257 | 366 | 224 | 498 | 325 | 427 | 290 | 346 | 200 | 366 | 305 |

Estimates are given for the real masters, grand master, and four synthetic cases with and without the *RM* constraint. Two cases are presented: the REB event has to have more than 2 and more than 3 stations with arrivals within 4 s from those in the XSELs.

Figure captions

Figure 1. Representations of the 45° dip slip fault (upper row) and δ=45°, θ=0° and λ=90° thrust (upper row). Left: focal mechanism; Central: style of faulting; Right: Radiation pattern. Lower row: δ=45°, θ=90° and λ=90° source mechanism. The style of faulting representation is determined by the strike direction, θ, of the fault clockwise against geographic North, the dip angle, δ, of the fault plane against the surface of the Earth, and the slip angle, λ, of the hanging block (orange) against the footwall block (white). The P and T axes denote the local pressure and tension axes during faulting, respectively. U denotes the slip direction of the hanging block. δP, δT, and δB are the plunge angles of the P, T, and neutral axes against the horizontal, respectively.

Figure 2. Distance ordered synthetic seismograms at seven IMS array stations for STI mechanism (upper left), ST mechanism (upper right), Expl2D (lower left), and Expl1D (lower right). All original synthetics are convolved with instrument responses. These four sources were chosen to produce different cases of isotropic (or quasi-isotropic) radiation pattern. The waveforms confirm this feature demonstrating positive first arrival polarity (dilatation motion) at practically all stations (except CMAR for ST mechanism). The differences in waveforms for Expl1D and Expl2D are related to the crustal model.

Figure 3. Double-couple representation of the CMT event.

Figure 4. Upper panel: Real (top 7) and synthetic waveforms band-filtered between 0.6 Hz and 4.5 Hz. Lower panel: Smoothed spectra for real and synthetic seismograms for stations CMAR, MKAR, and SONA. The difference in levels for predicted and observed data at 1.0 Hz is larger than 20 dB for CMAR comparing to MKAR.

Figure 5. Left panel: Location of sixteen real master evens (open diamonds), SGM (black triangle), and the regular grid with ~1° spacing (black circles). Right panel: The grid used in the local association. For each of sixteen grid nodes, there are two rings of 6 and 12 points at distances 0.225° and 0.450°, respectively. For a given master, the LA creates 19 event hypotheses by reducing all arrival times to origin times using the relevant master/station travel times. The point with the smallest RMS origin time residual represents the location of the best event hypothesis.

Figure 6. Probability density functions of the average *CC, CC_AVE,* for all associated phases (left panel) and those in the XSEL after the conflict resolution (CR) without the relative amplitude limit (No RM) applied (right panel) for five studied cases. Counting in 0.1-wide bins starting from *CC_AVE*=0.1. Notice the lin-log scales.

Figure 7. Probability density functions of the cumulative *CC, CC_CUM,* for all associated phases (left panel) and those in the XSEL after the conflict resolution (CR) without the relative amplitude limit (NoRM) applied (right panel) for five synthetic masters.

Figure 8. Probability density functions of the relative magnitude*, RM,* for all associated phases (left panel) and those in the XSEL after the conflict resolution (CR) without the relative amplitude limit (NoRM) applied (right panel) for five synthetic masters.

Figure 9. Comparison of the probability density functions of *CC* estimated using five synthetic masters for all associated arrivals (left panel) and those in the relevant XSEL (right panel).

Figure 10. Left panel: Comparison of the probability density functions of *SNR_CC* estimated using five synthetic masters for all associated arrivals. Right panel: Probability density functions of *SNR_CC* estimated for two versions of conflict resolution (with and without d*RM*) with the Expl1D synthetic master.

Figure 11. Comparison of the PDFs of d*RM* estimated using five synthetic masters for the relevant XSELs with (left panel) and without (right panel) d*RM* threshold.

Figure 12. Comparison of three real and two synthetic master sets.

Figure 13. Probability density functions of the deviation of relative magnitude from the relevant network average.

Figure 14. Frequency distributions of the residual azimuth, d*AZ*, for seven IMS stations as obtained from the real and synthetic master events.

Figure 15. Frequency distributions of the residual slowness, d*SLO*, for seven IMS stations as obtained from the real and synthetic master events.

Figure 16. Real (upper) and synthetic (lower) seismograms filtered between 0.5Hz and 2Hz for source depth 1.2 km.

Figures

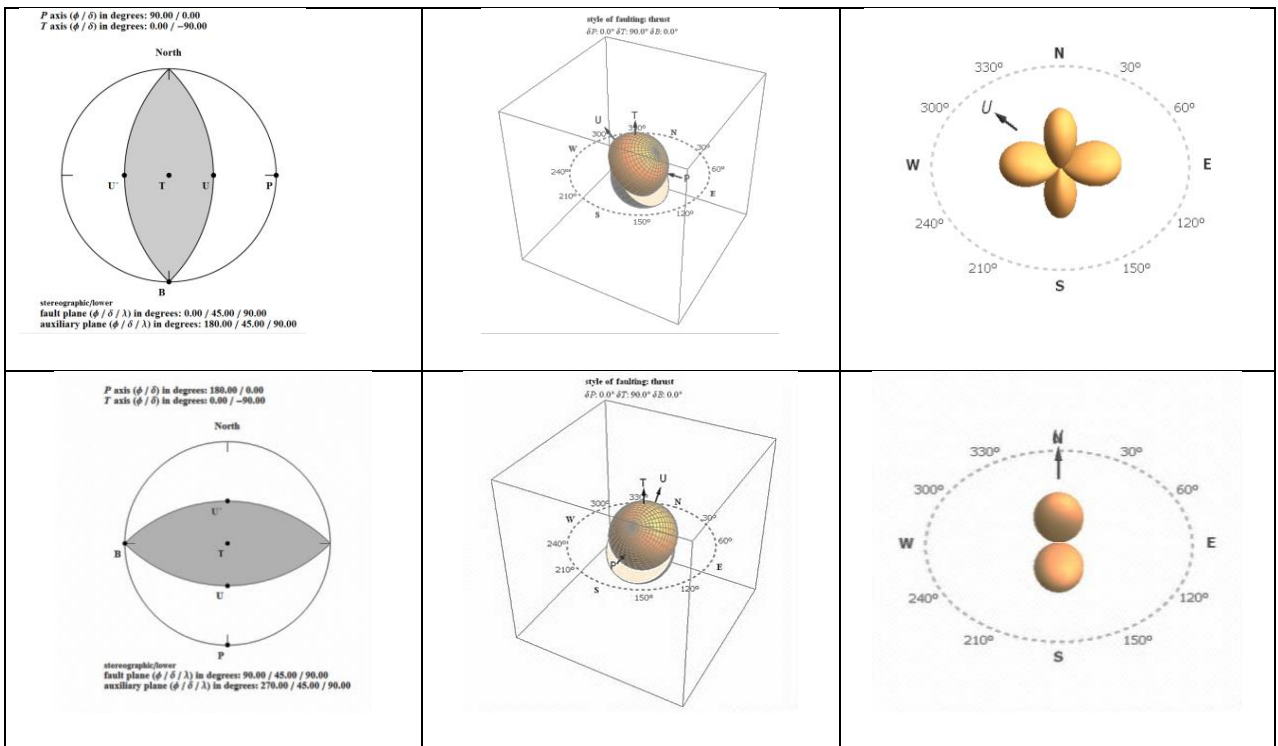

Figure 1. Representations of the 45° dip slip fault (upper row) and δ=45°, θ=0° and λ=90° thrust (upper row). Left: focal mechanism; Central: style of faulting; Right: Radiation pattern. Lower row: δ=45°, θ=90° and λ=90° source mechanism. The style of faulting representation is determined by the strike direction, θ, of the fault clockwise against geographic North, the dip angle, δ, of the fault plane against the surface of the Earth, and the slip angle, λ, of the hanging block (orange) against the footwall block (white). The P and T axes denote the local pressure and tension axes during faulting, respectively. U denotes the slip direction of the hanging block. δP, δT, and δB are the plunge angles of the P, T, and neutral axes against the horizontal, respectively.

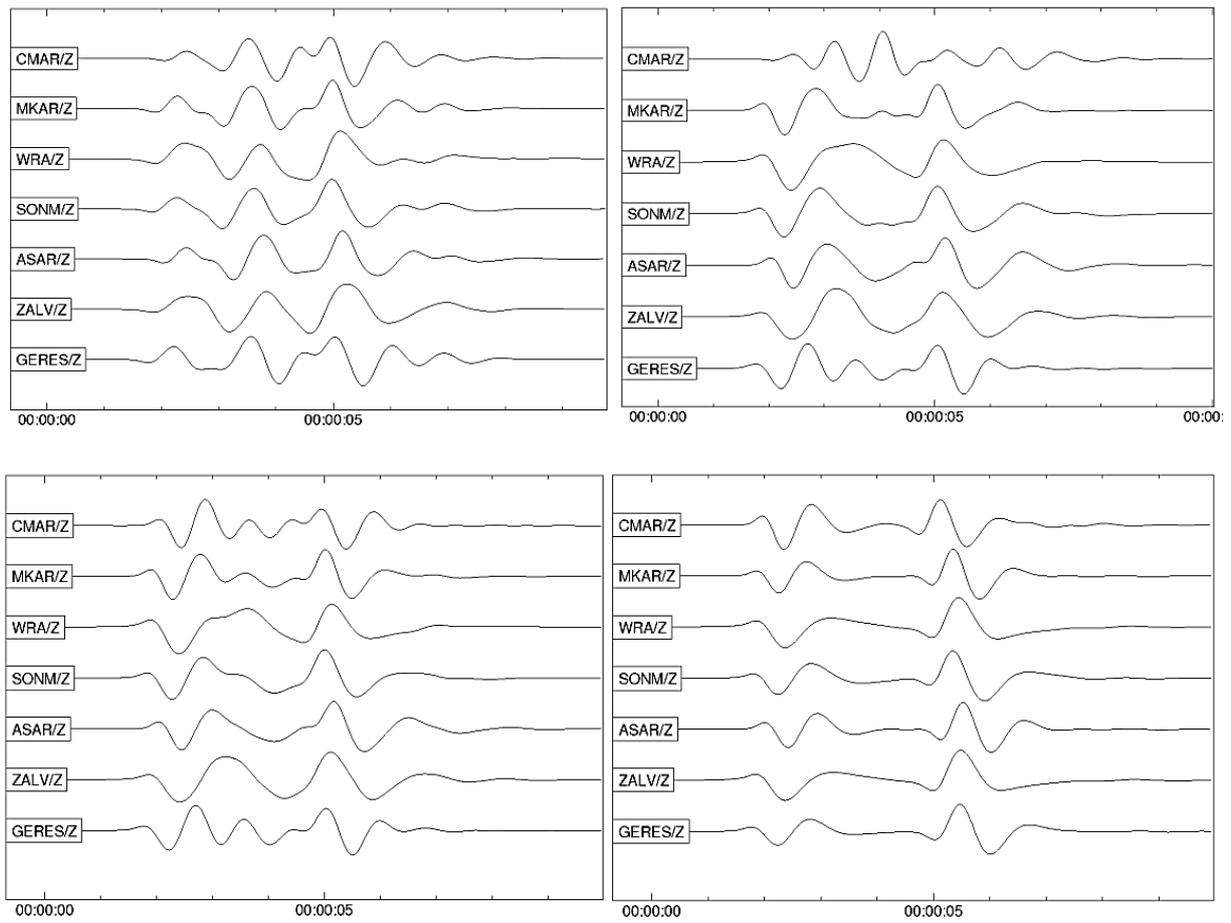

Figure 2. Distance ordered synthetic seismograms at seven IMS array stations for STI mechanism (upper left), ST mechanism (upper right), Expl2D (lower left), and Expl1D (lower right). All original synthetics are convolved with instrument responses. These four sources were chosen to produce different cases of isotropic (or quasi-isotropic) radiation pattern. The waveforms confirm this feature demonstrating positive first arrival polarity (dilatation motion) at practically all stations (except CMAR for ST mechanism). The differences in waveforms for Expl1D and Expl2D are related to the crustal model.

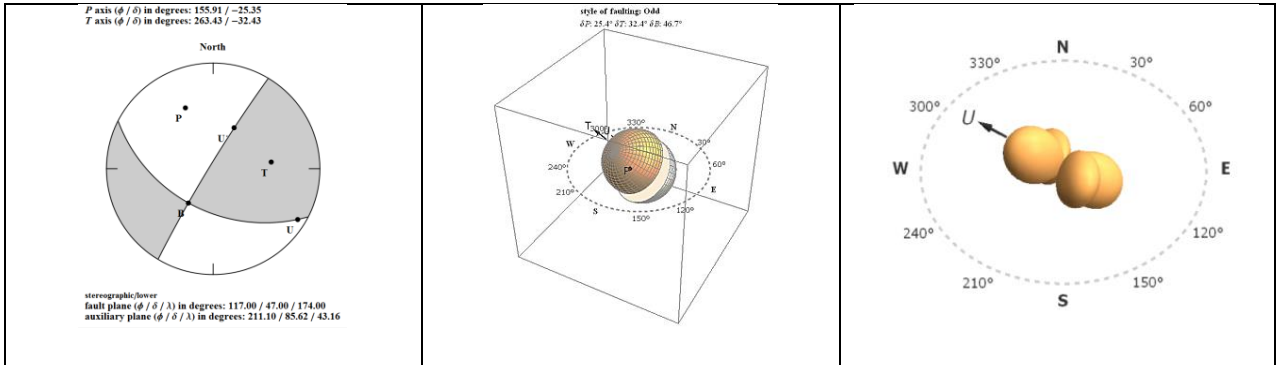
Figure 3. Double-couple representation of the CMT event.

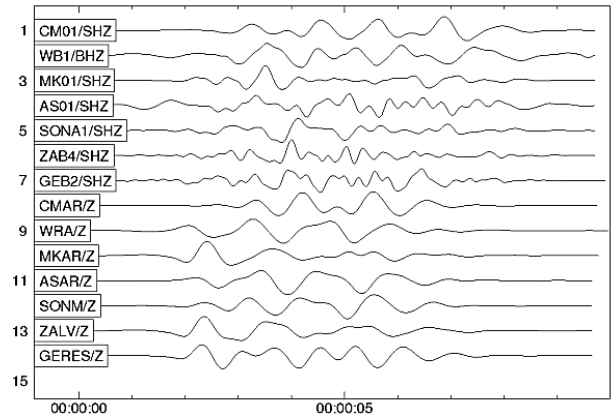

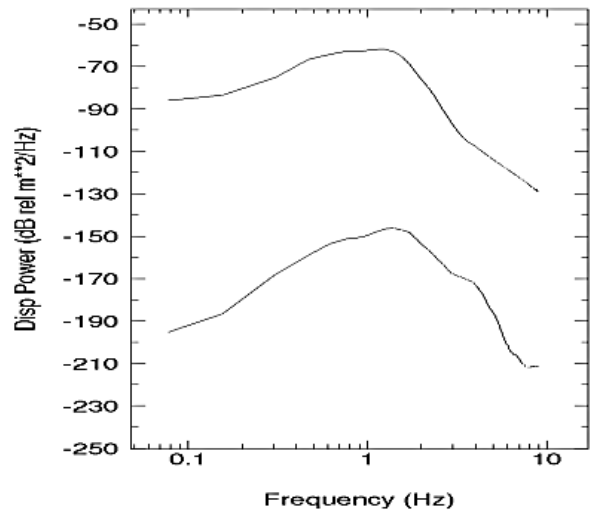

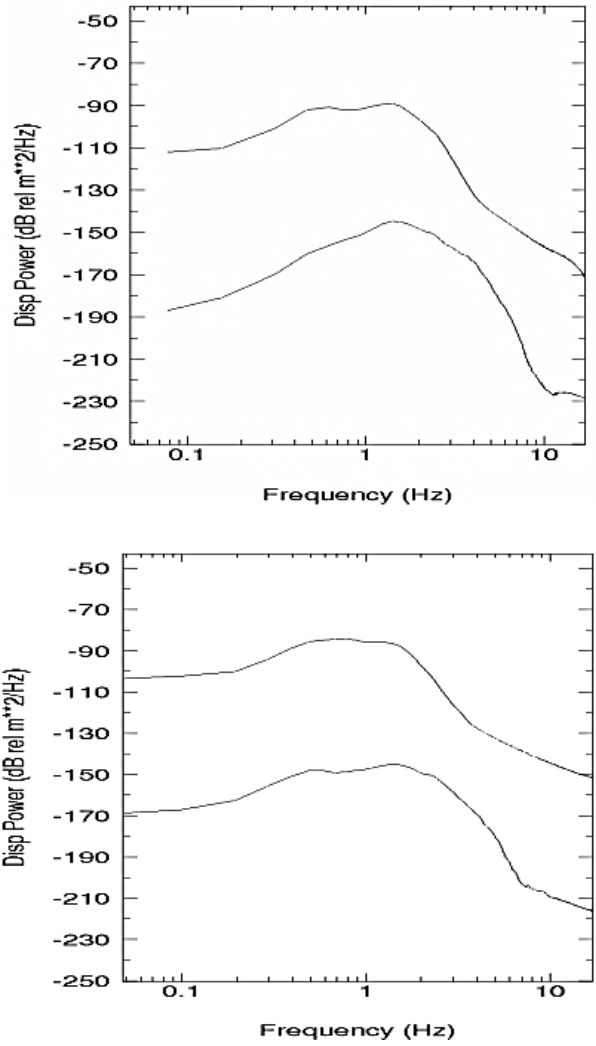

Figure 4. Upper panel: Real (top 7) and synthetic waveforms band-filtered between 0.6 Hz and 4.5 Hz. Lower panel: Smoothed spectra for real and synthetic seismograms for stations CMAR, MKAR, and SONA. The difference in levels for predicted and observed data at 1.0 Hz is larger than 20 dB for CMAR comparing to MKAR.

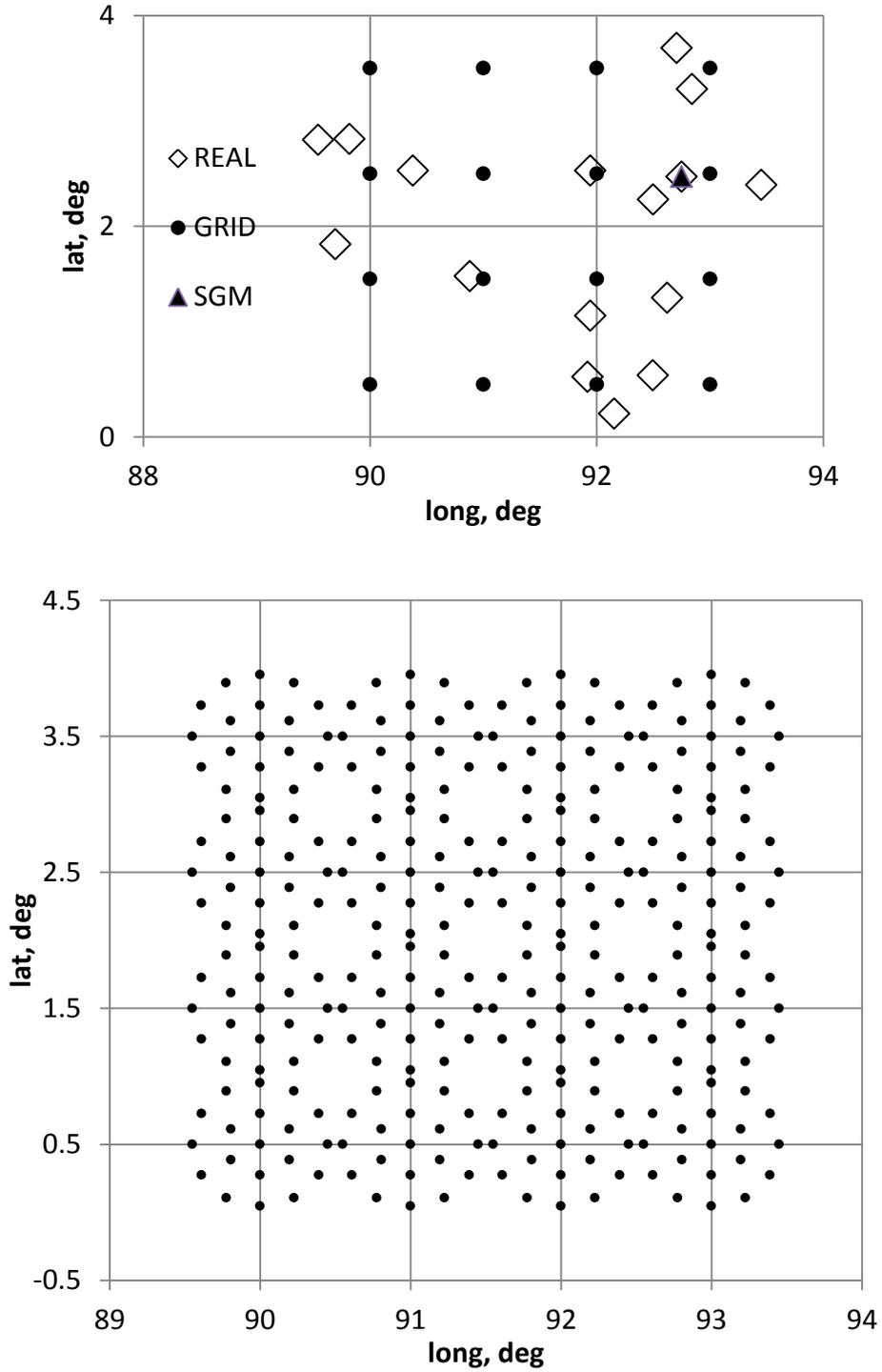

Figure 5. Left panel: Location of sixteen real master evens (open diamonds), SGM (black triangle), and the regular grid with ~1° spacing (black circles). Right panel: The grid used in the local association. For each of sixteen grid nodes, there are two rings of 6 and 12 points at distances 0.225° and 0.450°, respectively. For a given master, the LA creates 19 event hypotheses by reducing all arrival times to origin times using the relevant master/station travel

times. The point with the smallest RMS origin time residual represents the location of the best event hypothesis.

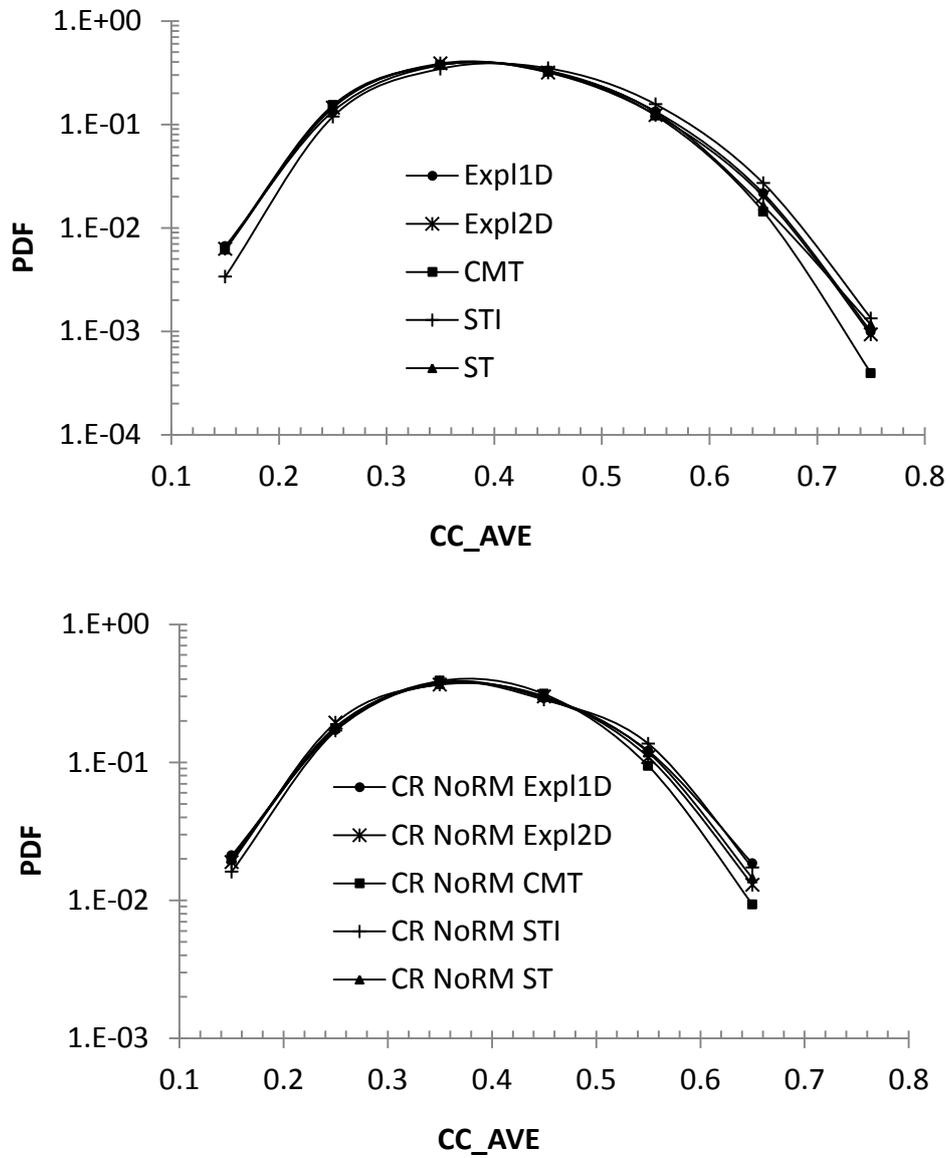

Figure 6. Probability density functions of the average *CC, CC_AVE,* for all associated phases (left panel) and those in the XSEL after the conflict resolution (CR) without the relative amplitude limit (No RM) applied (right panel) for five studied cases. Counting in 0.1-wide bins starting from *CC_AVE*=0.1. Notice the lin-log scales.

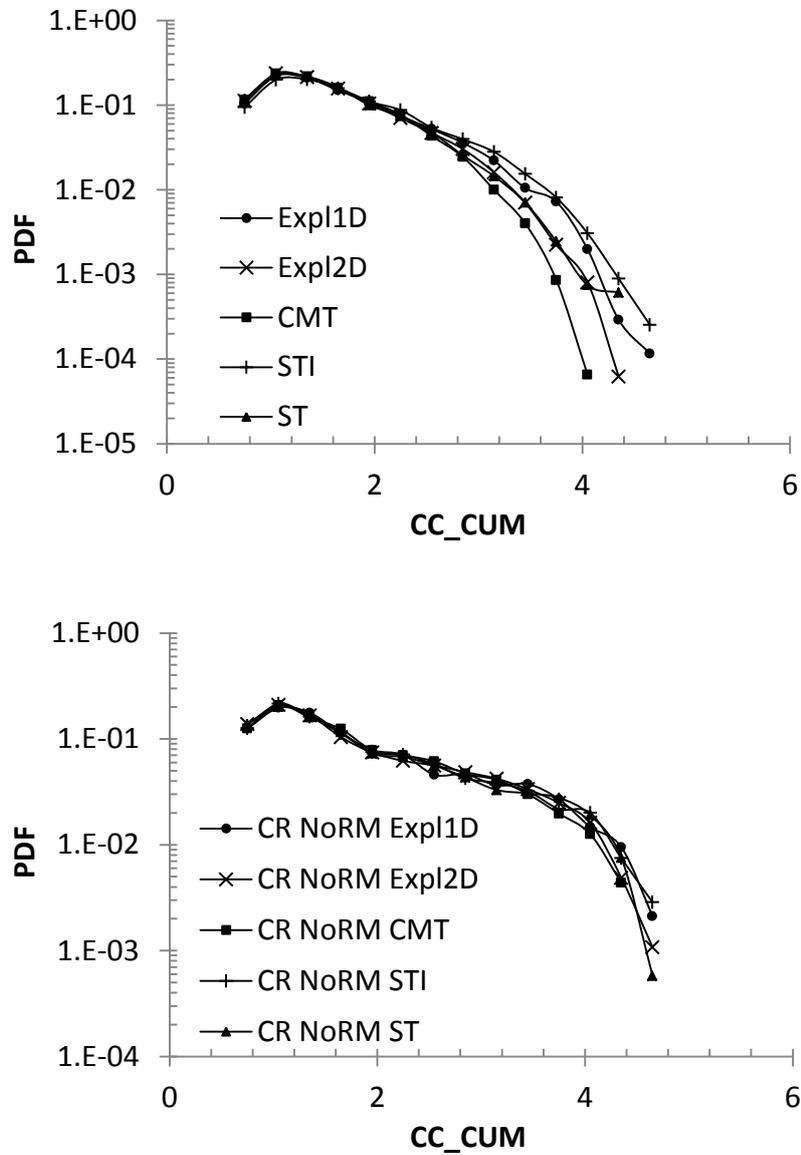

Figure 7. Probability density functions of the cumulative *CC, CC_CUM,* for all associated phases (left panel) and those in the XSEL after the conflict resolution (CR) without the relative amplitude limit (NoRM) applied (right panel) for five synthetic masters.

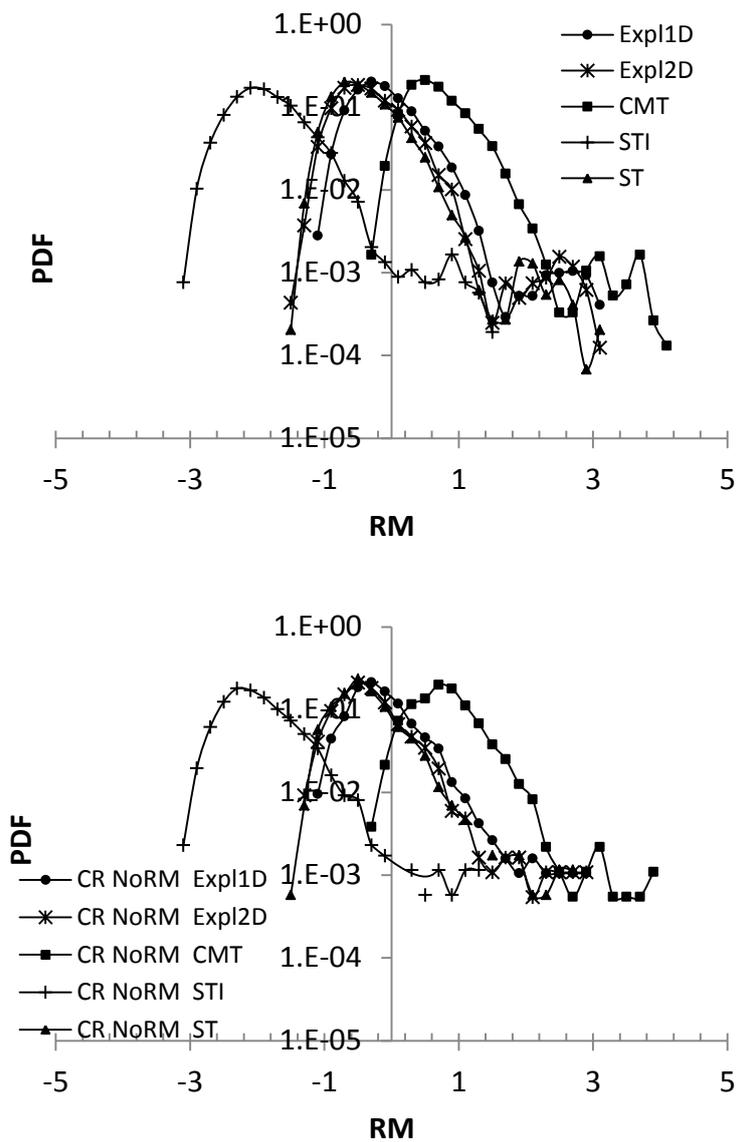

Figure 8. Probability density functions of the relative magnitude, *RM,* for all associated phases (left panel) and those in the XSEL after the conflict resolution (CR) without the relative amplitude limit (NoRM) applied (right panel) for five synthetic masters.

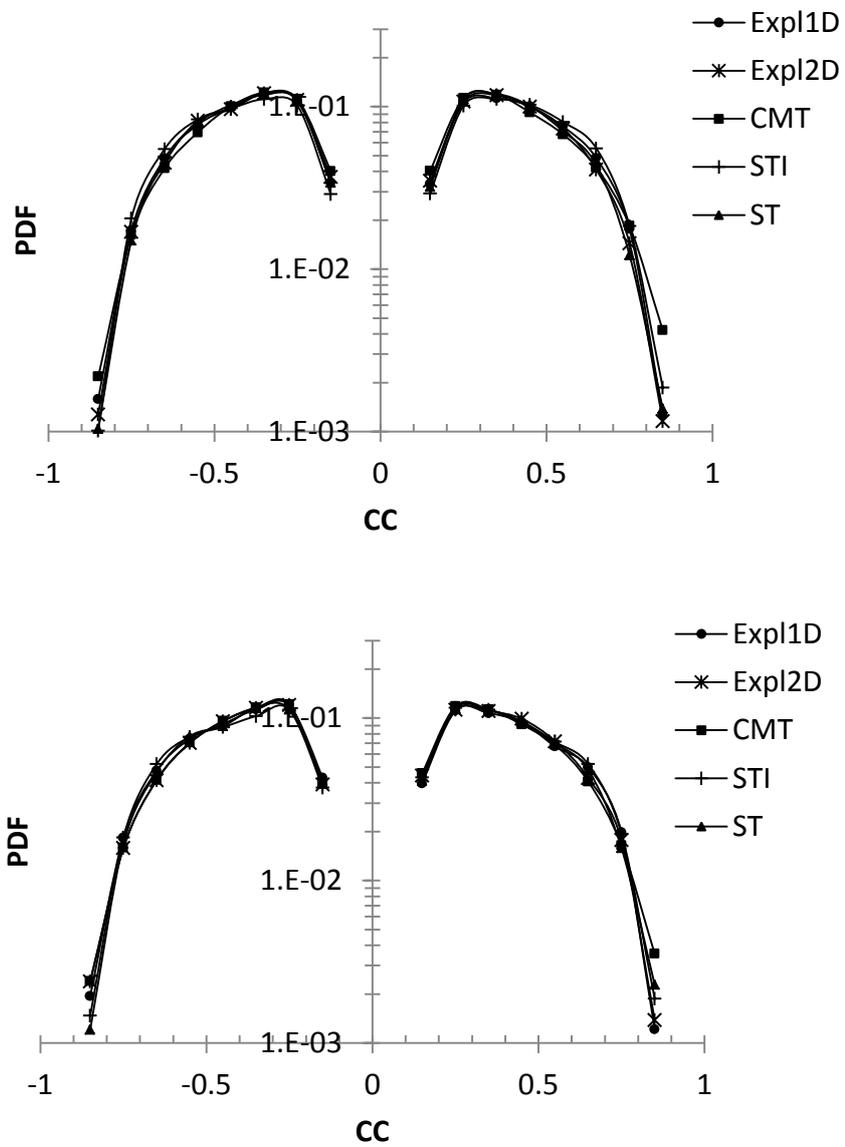

Figure 9. Comparison of the probability density functions of *CC* estimated using five synthetic masters for all associated arrivals (left panel) and those in the relevant XSEL (right panel).

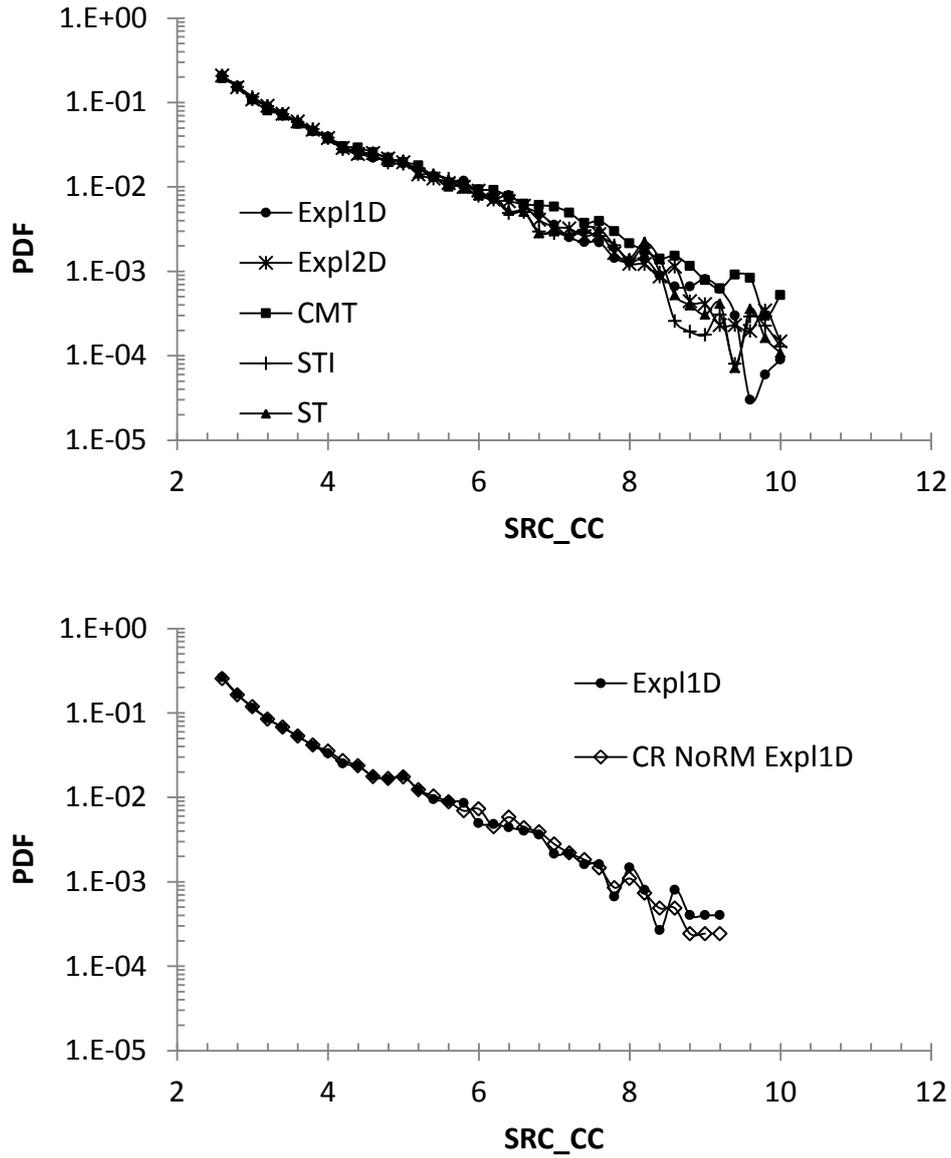

Figure 10. Left panel: Comparison of the probability density functions of *SNR_CC* estimated using five synthetic masters for all associated arrivals. Right panel: Probability density functions of *SNR_CC* estimated for two versions of conflict resolution (with and without d*RM*) with the Expl1D synthetic master.

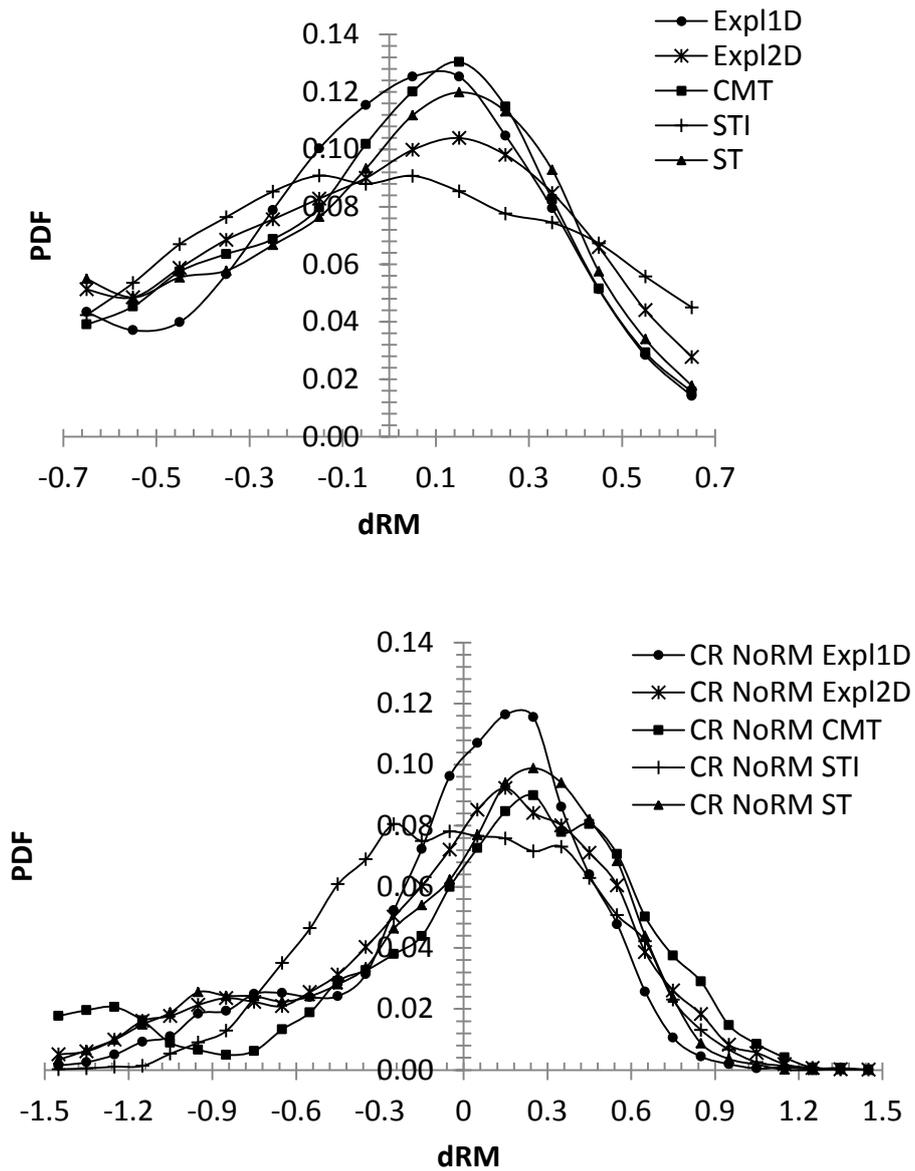

Figure 11. Comparison of the PDFs of d*RM* estimated using five synthetic masters for the relevant XSELs with (left panel) and without (right panel) d*RM* threshold.

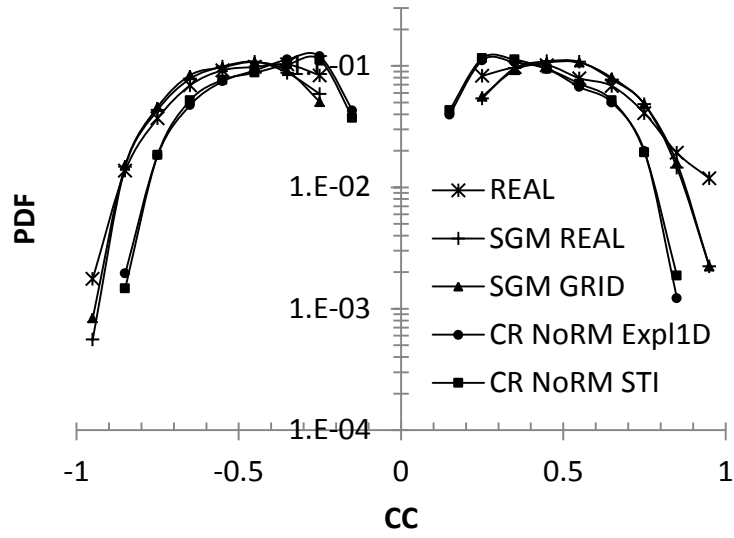

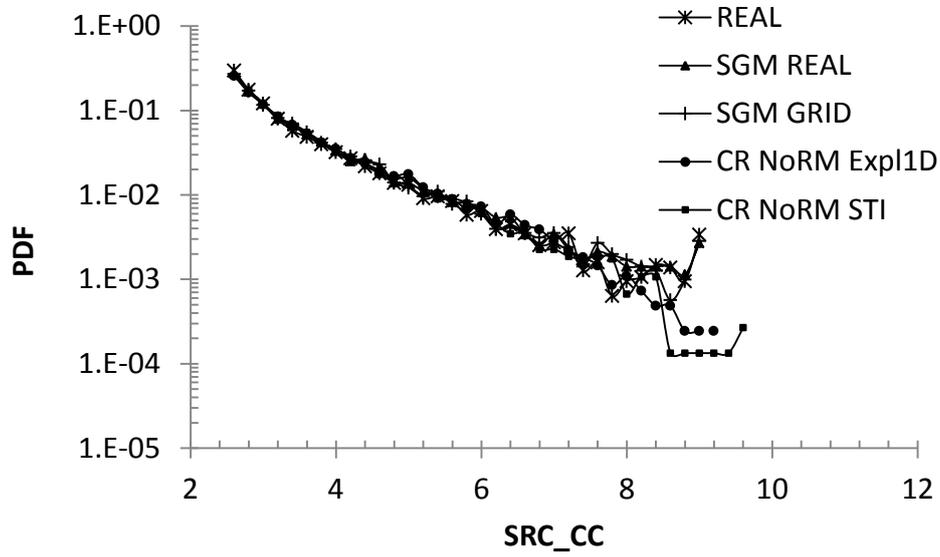

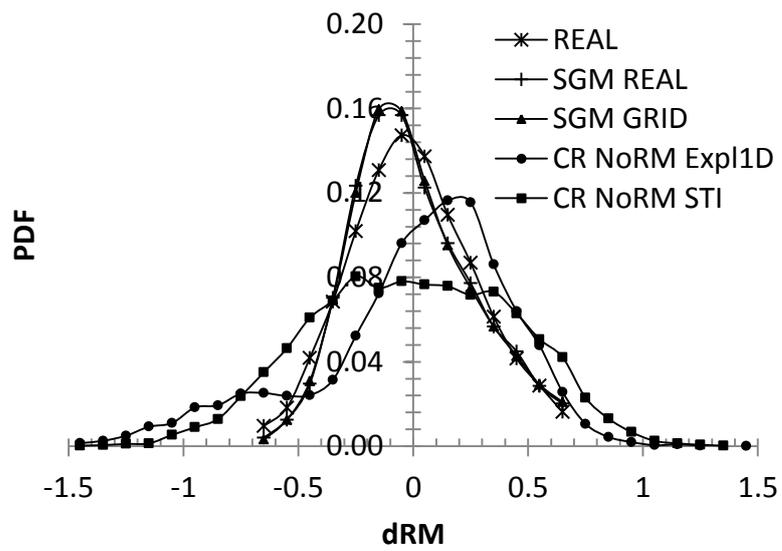

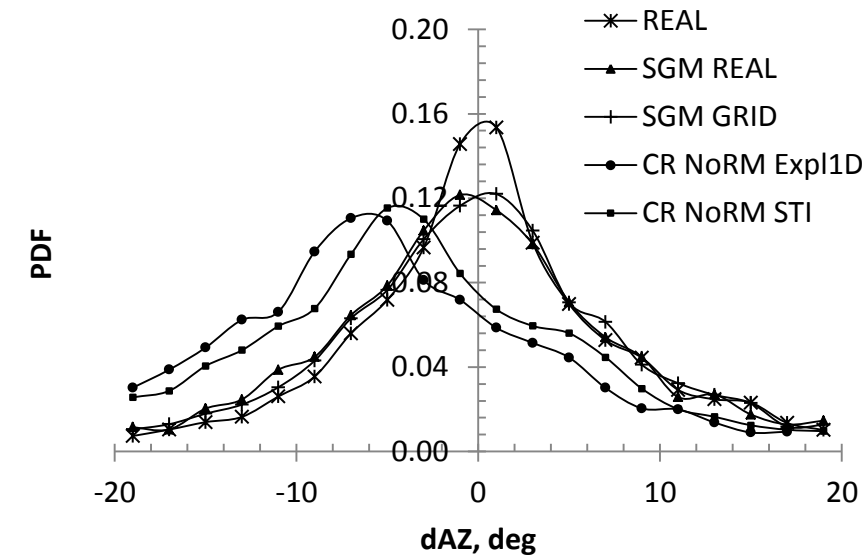

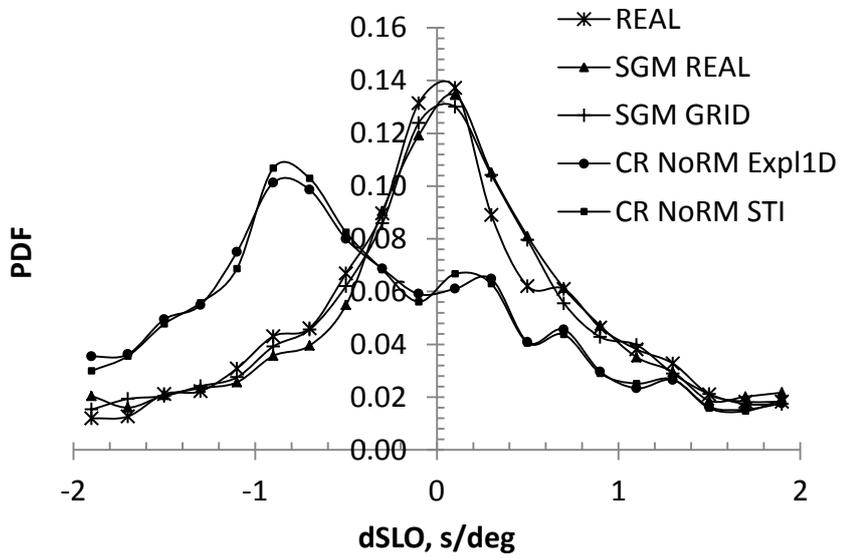

Figure 12. Comparison of three real and two synthetic master sets.

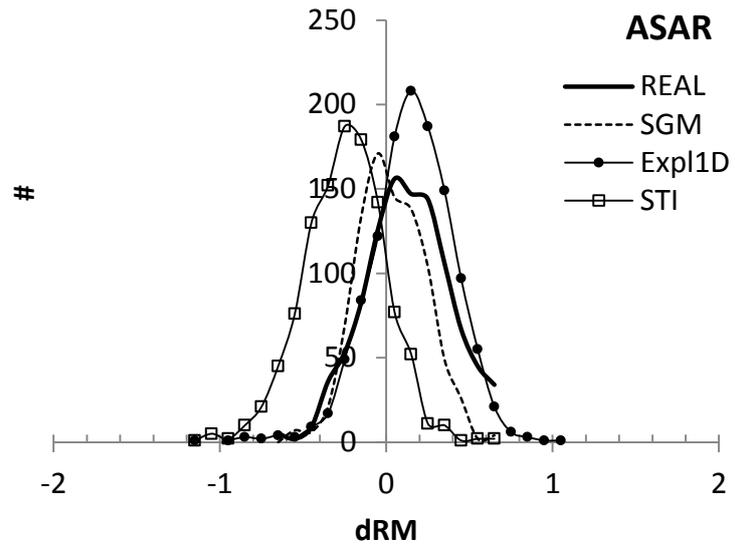
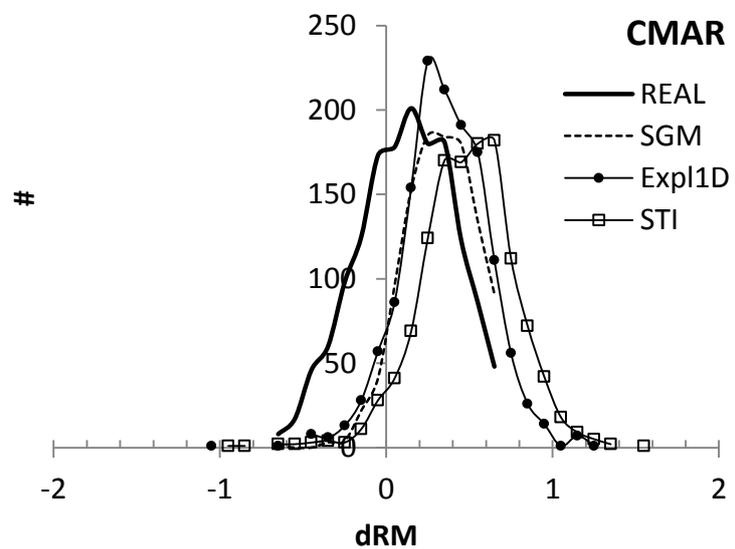
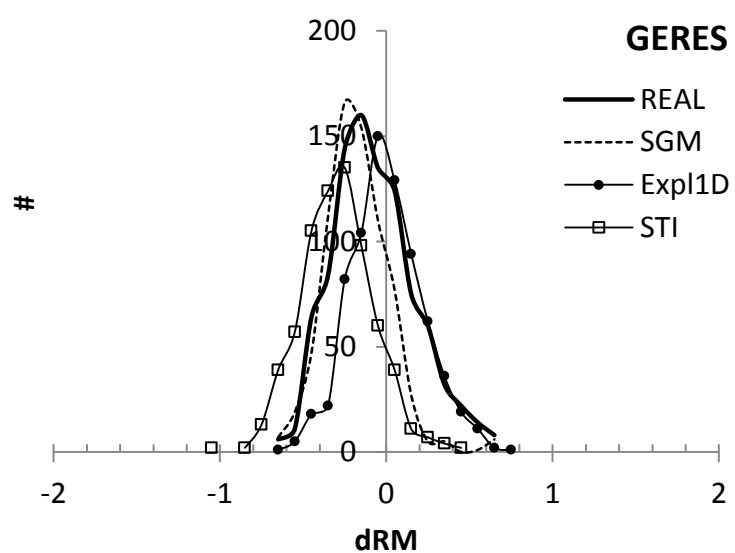

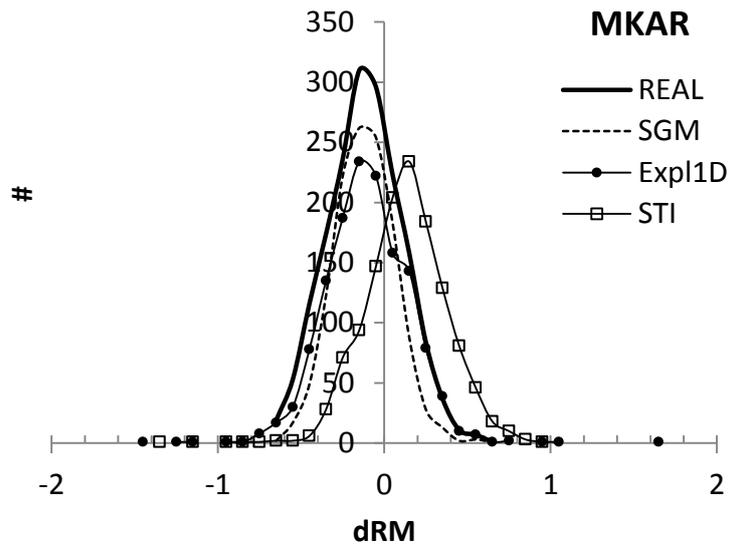
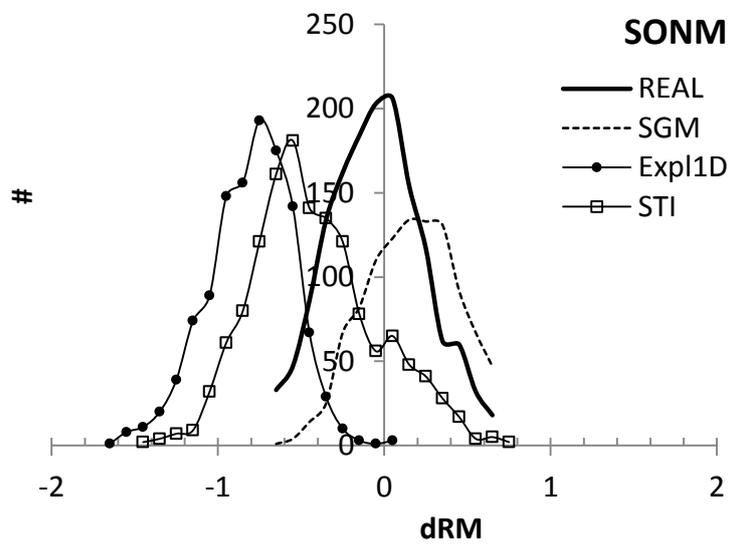

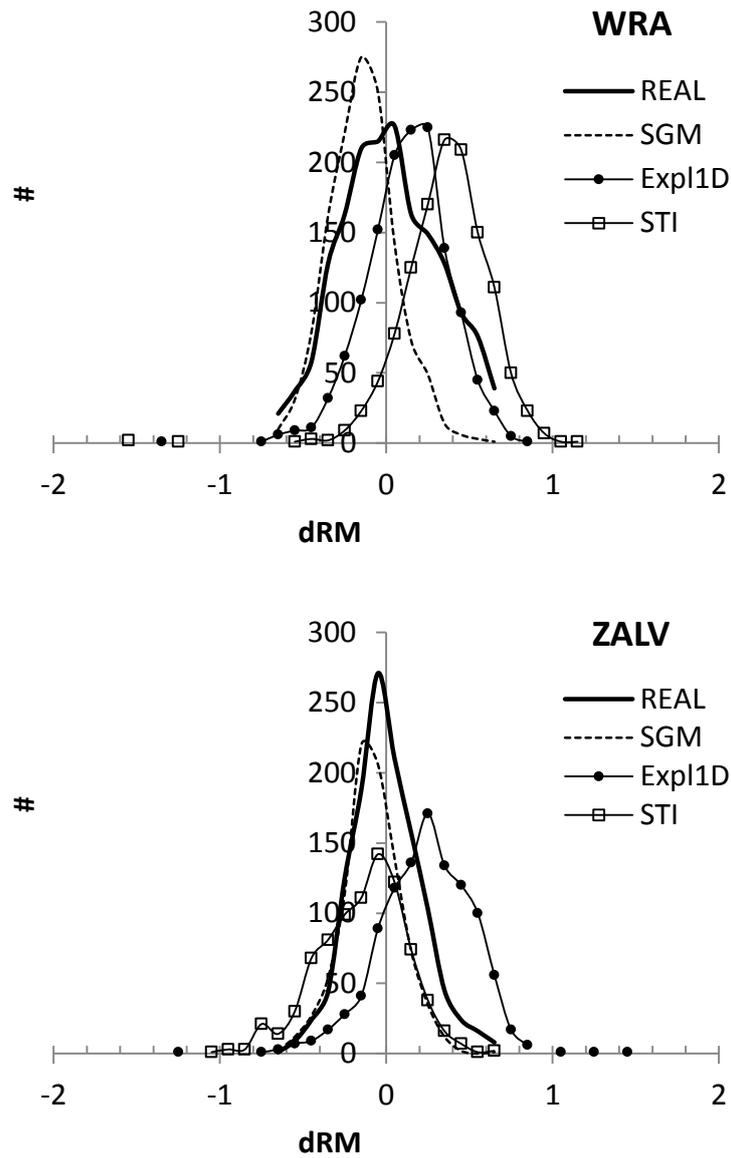

Figure 13. Probability density functions of the deviation of relative magnitude from the relevant network average.

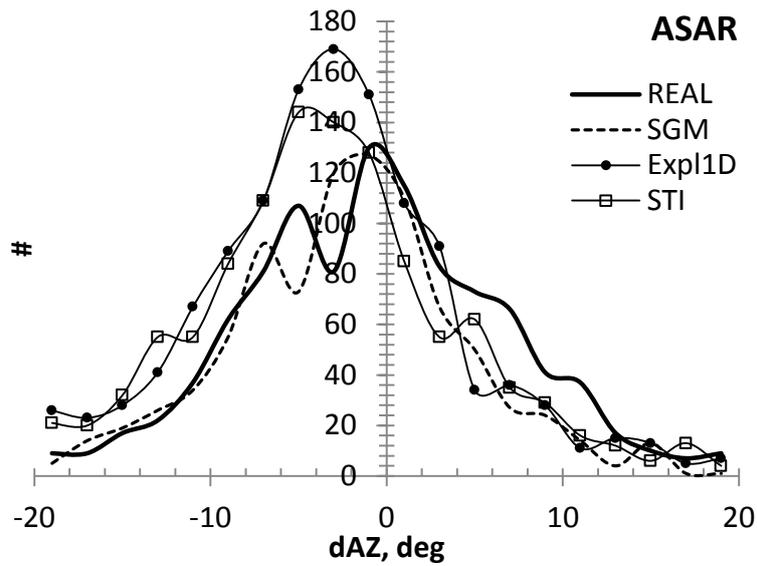

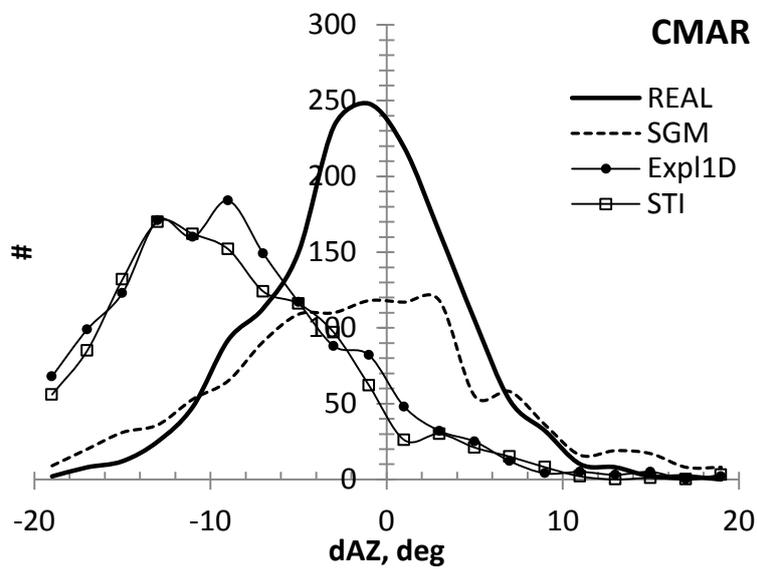

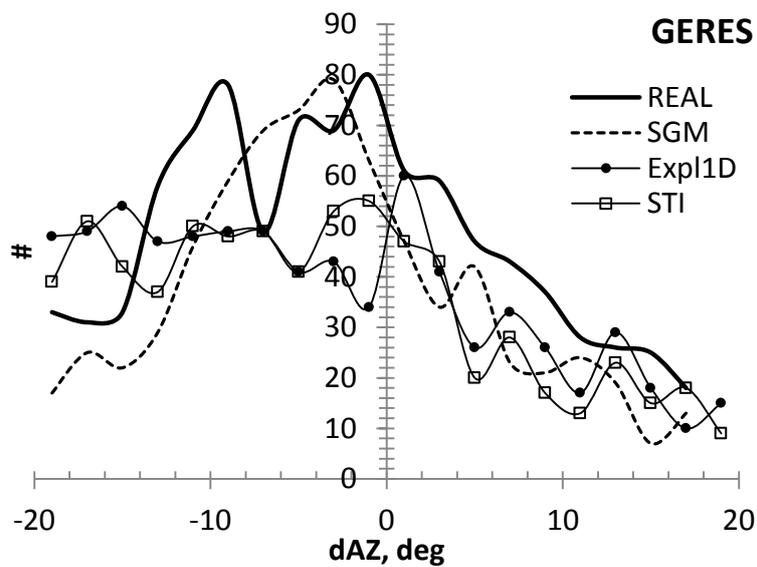

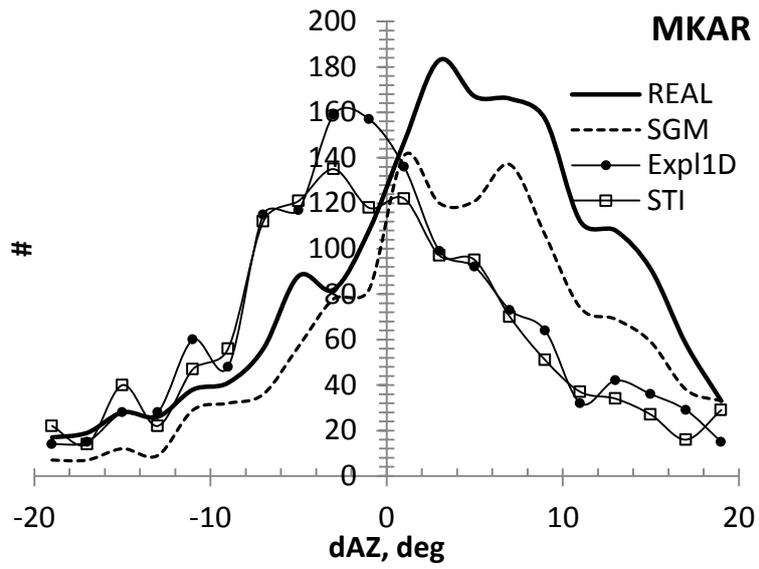

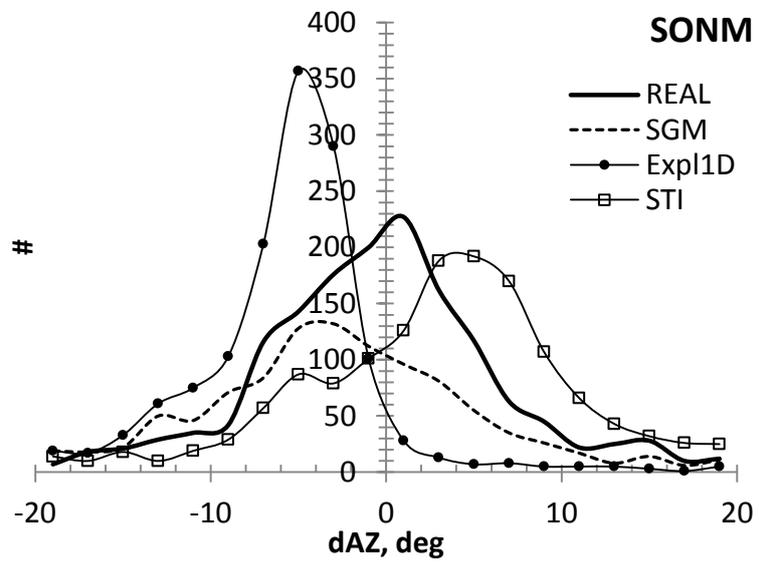

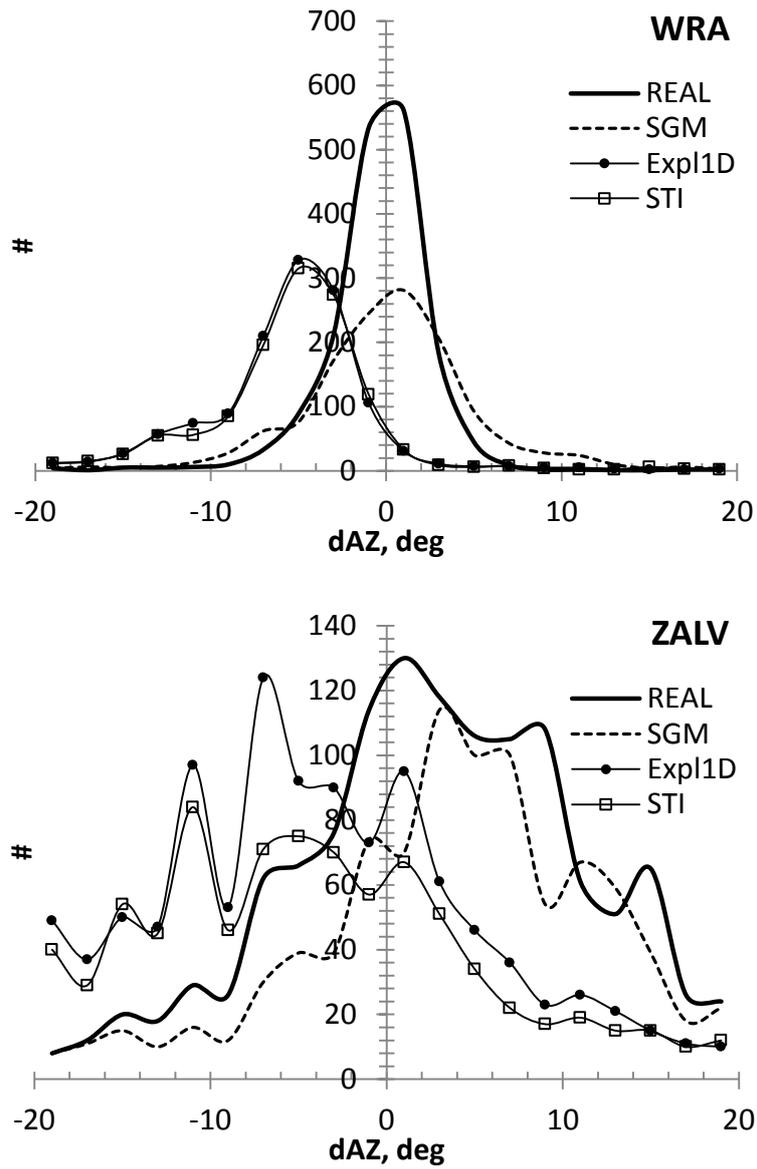

Figure 14. Frequency distributions of the residual azimuth, d*AZ*, for seven IMS stations as obtained from the real and synthetic master events.

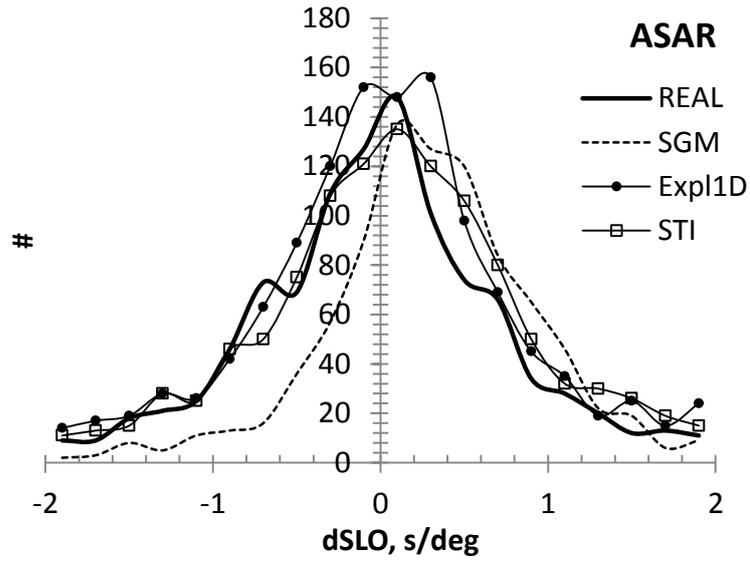

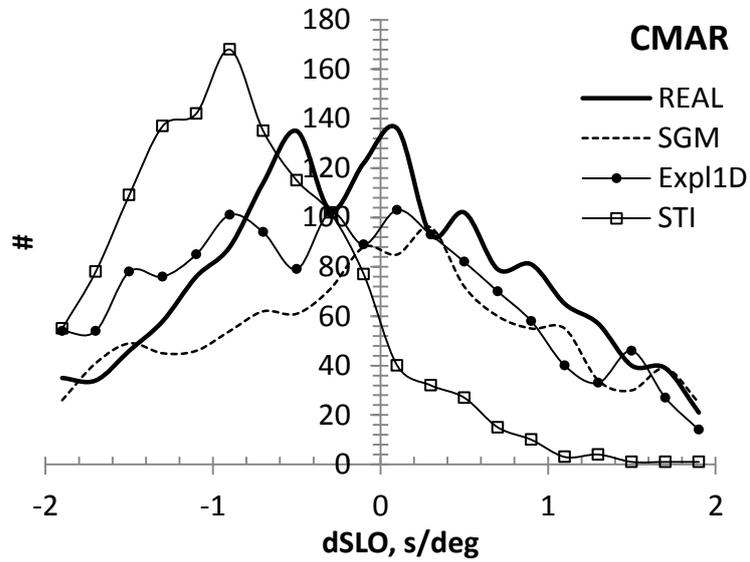

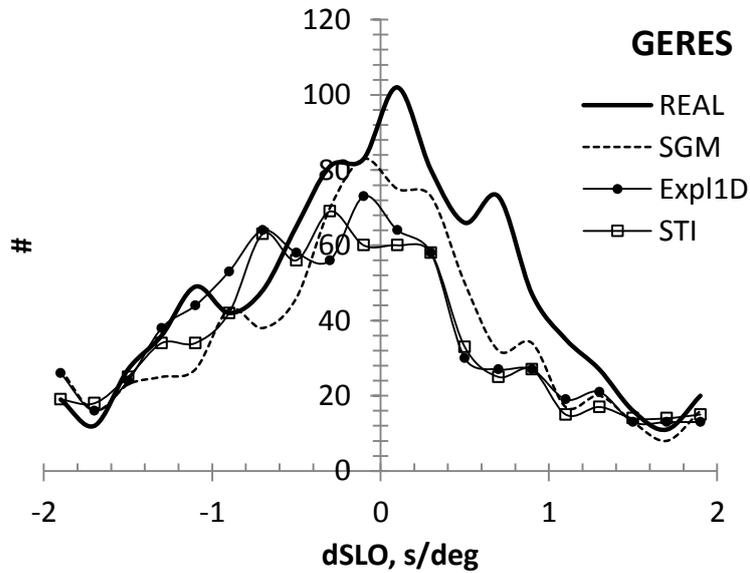

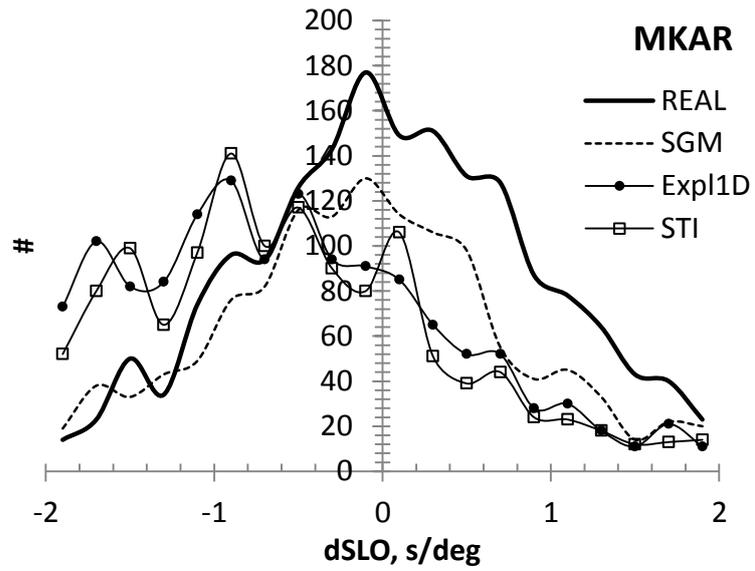
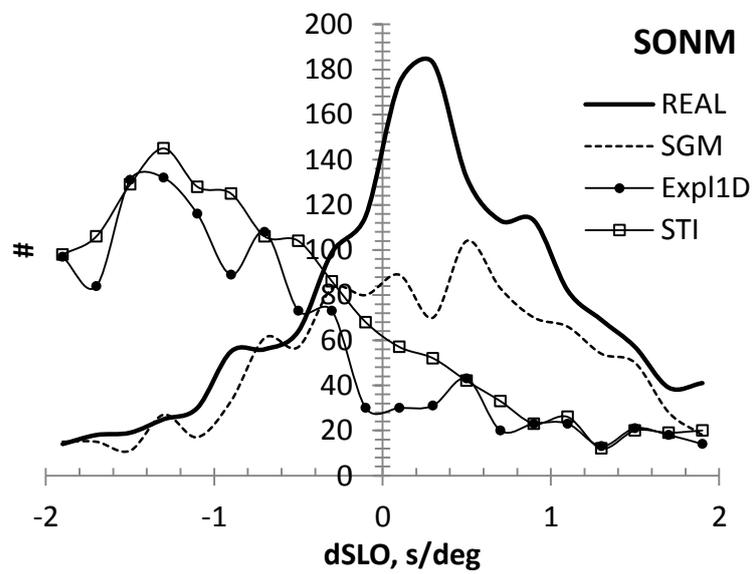

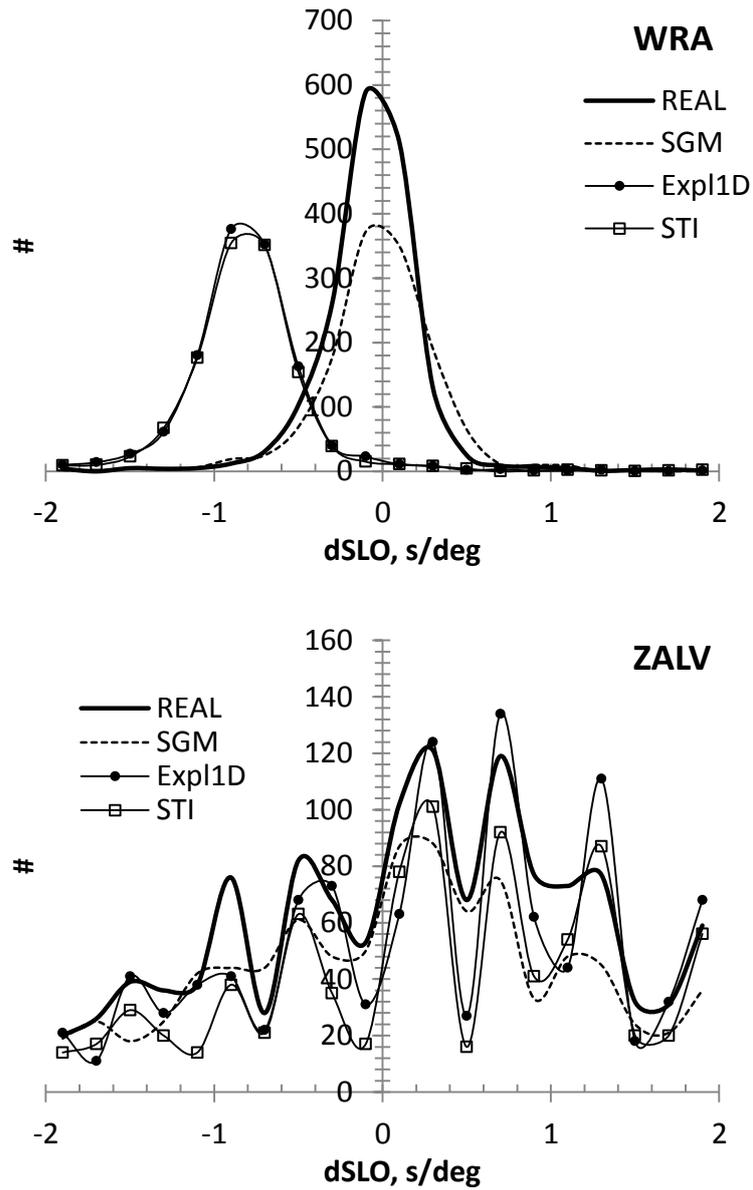

Figure 15. Frequency distributions of the residual slowness, d*SLO*, for seven IMS stations as obtained from the real and synthetic master events.

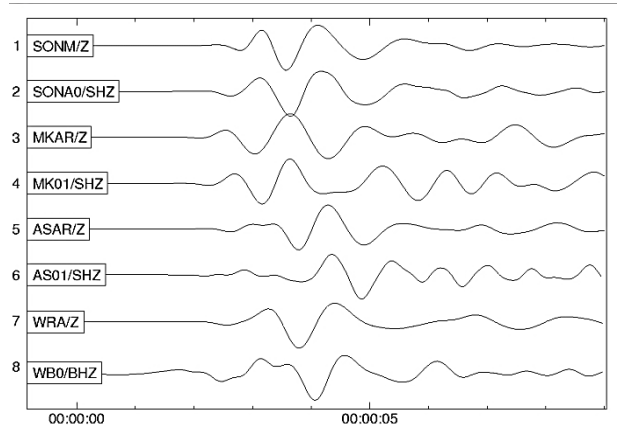

Figure 16. Real (upper) and synthetic (lower) seismograms filtered between 0.5Hz and 2Hz for source depth 1.2 km.